\documentclass[preprint,12pt]{elsarticle}

\usepackage{amssymb}
\usepackage{amsmath}

\usepackage{makecell}
\usepackage{fancyvrb}
\usepackage{graphicx}
\usepackage{multirow}
\usepackage{array}
\usepackage{caption}
\usepackage{subcaption}
\usepackage{listings}
\usepackage{booktabs}
\usepackage{url}
\usepackage{rotating}
\usepackage{xcolor}

\lstdefinelanguage{yaml}{
	frame=single,
	showstringspaces=false,
	tabsize=2,
	breaklines=true,
	captionpos=b,
	lineskip=-5pt,
	keywords={true,false,null,y,n},
	keywordstyle=\color{blue}\bfseries,
	basicstyle=\ttfamily\scriptsize,
	comment=[l]{\#},
	commentstyle=\color{gray}\ttfamily,
	stringstyle=\color{red}\ttfamily,
	moredelim=[l]{:},
	moredelim=[s]{[}{]},
	moredelim=[s]{\{}{\}},
	numbers=left,
	sensitive=true
}

\lstdefinelanguage{logs}{
	frame=single,
	showstringspaces=false,
	tabsize=2,
	breaklines=true,
	captionpos=b,
	lineskip=-5pt,
	basicstyle=\ttfamily\scriptsize,
	numbers=left,
	sensitive=true,
	moredelim=*[is][\color{blue}]{<!*}{*!>},
}

\journal{Computers \& Security}

\begin{document}

\begin{frontmatter}



\title{CAM-LDS: Cyber Attack Manifestations for Automatic Interpretation of System Logs and Security Alerts}


\author{Max Landauer, Wolfgang Hotwagner, Thorina Boenke, Florian Skopik, Markus Wurzenberger} 

\affiliation{organization={Austrian Institute of Technology},
            addressline={Center for Digital Safety \& Security}, 
            city={Vienna},
            postcode={1210},
            country={Austria}}

\begin{abstract}
Log data are essential for intrusion detection and forensic investigations. However, manual log analysis is tedious due to high data volumes, heterogeneous event formats, and unstructured messages. Even though many automated methods for log analysis exist, they usually still rely on domain-specific configurations such as expert-defined detection rules, handcrafted log parsers, or manual feature-engineering. Crucially, the level of automation of conventional methods is limited due to their inability to semantically understand logs and explain their underlying causes. In contrast, Large Language Models enable domain- and format-agnostic interpretation of system logs and security alerts. Unfortunately, research on this topic remains challenging, because publicly available and labeled data sets covering a broad range of attack techniques are scarce. To address this gap, we introduce the Cyber Attack Manifestation Log Data Set (CAM-LDS), comprising seven attack scenarios that cover 81 distinct techniques across 13 tactics and collected from 18 distinct sources within a fully open-source and reproducible test environment. We extract log events that directly result from attack executions to facilitate analysis of manifestations concerning command observability, event frequencies, performance metrics, and intrusion detection alerts. We further present an illustrative case study utilizing an LLM to process the CAM-LDS. The results indicate that correct attack techniques are predicted perfectly for approximately one third of attack steps and adequately for another third, highlighting the potential of LLM-based log interpretation and utility of our data set.
\end{abstract}


\begin{highlights}
\item Public labeled log data sets of attack traces and artifacts.
\item Analysis and categorization of cyber attack manifestations.
\item LLM-based interpretation of system logs and security alerts.
\end{highlights}

\begin{keyword}
system log data \sep cyber attack data set \sep log interpretation \sep large language models \sep intrusion detection \sep security alerts



\end{keyword}

\end{frontmatter}



\section{Introduction} \label{intro}

Protection against the ever-growing increase of cyber attacks is of paramount importance for organizations across all sectors. Among the various domains of cyber security, log data analysis stands out as both highly valuable and inherently complex. On the one hand, system log data presents a persistent and fine-grained record of activities occurring on networks and computer systems. Accordingly, they constitute a fundamental source of information for the detection of immanent attacks as well as forensic investigations of malicious or undesired activities after the fact \cite{alzu2025cyberattack}.

On the other hand, log data appear in large volumes, involve heterogeneous formats, and often contain unstructured messages. Their interpretation therefore requires substantial manual effort and expert knowledge \cite{qi2023loggpt, zhang2025llm, shanto2024console}. Although automated analysis methods exist, their practical applicability remains limited: signature-based detection approaches rely on manually crafted rules that are difficult to maintain and ineffective against novel and unknown attacks \cite{khraisat2019survey}; anomaly-based methods require large amounts of representative training data \cite{liu2024logprompt}, frequently suffer from high false positive rates \cite{khraisat2019survey}, depend on manual feature engineering \cite{palma2025leveraging}, often struggle to adapt to evolving attack patterns \cite{palma2025leveraging}, and typically report statistical deviations without providing semantic explanations of the underlying causes \cite{hans2025security, liu2024logprompt, zhang2025llm}.

The emergence of Large Language Models (LLM) has significantly changed how security engineers approach log analysis. Recent studies in Security Operations Centers (SOC) indicate that practitioners increasingly use LLMs to interpret and explain system logs and other low-level telemetry data \cite{singh2025llms}. Thereby, the key capability of LLMs is their ability to semantically analyze log data in a manner analogous to natural language \cite{karlsen2024benchmarking}, which supports engineers in understanding unfamiliar log formats or unknown events without manual lookup of documentations \cite{liu2024interpretable}.

Aside from direct interpretation of log snippets, LLMs have been successfully demonstrated for various log analysis tasks such as log parsing, failure detection, intrusion detection, root cause analysis, and summarization \cite{mudgal2023assessment, liu2025loglm, balogh2024using}. In contrast to traditional approaches that require separate specialized models for different tasks and log formats, LLMs provide a unified framework capable of supporting or autonomously performing multiple analysis objectives \cite{liu2025loglm, karlsen2024benchmarking}. Consequently, the automation of SOC workflows through LLM-based methods is expected to expand further, even though several challenges remain, including significant computational overhead, high costs, hallucinations, inconsistent outputs, ambiguous terminology, oversimplified reasoning, and limited transparency \cite{habibzadeh2025large}.

A major obstacle for advancing research on LLM-based log interpretation is the lack of suitable public log data sets for evaluation \cite{habibzadeh2025large}. Researchers therefore often resort to private data sets \cite{shanto2024console, cotti2025ontologx}, which hinder reproducibility and comparability, or data sets without security context \cite{liu2024interpretable, liu2025loglm, karlsen2024benchmarking}, which limits the relevance of results for security applications. While some log data sets from the intrusion detection domain exist, they typically focus on network traffic \cite{sharafaldin2018toward, bagui2023introducing, elam2025introducing}, are restricted to Windows environments \cite{mordor, evtxas, evtxtoma, atomic}, or originate from noisy environments in which isolating and labeling attack traces is challenging \cite{perez2025comiset, landauer2022maintainable}. Critically, most security-related data sets cover only a limited set of attack tactics and techniques, thereby failing to represent the diversity of real-world kill chains and limiting the scope of evaluations.

To address these challenges and enable reproducible research in log interpretation and related areas, we introduce the \textit{Cyber Attack Manifestation Log Data Set (CAM-LDS)}, a collection of system log data and security alerts that are direct consequences of cyber attacks. In contrast to prior data set collection strategies, we specifically focus on implementing a broad range of distinct attack techniques and arranging them into multiple coherent kill chains. Thereby, we pursue realistic attack execution within an idle network, allowing us to collect labeled manifestations of attacker behavior while minimizing unrelated background noise. To the best of our knowledge, CAM-LDS is the first publicly available data set specifically designed to support research on LLM-based interpretation of system logs and security alerts. 

We release the system log data\footnote{\url{https://zenodo.org/records/18390560}} and network packet captures\footnote{\url{https://zenodo.org/records/18701094}} online. In addition, we provide the virtual testbed infrastructure \textit{AttackBed} together with all implemented attack automation scripts as open-source code\footnote{\url{https://github.com/ait-testbed/attackbed}}. Furthermore, we publish the artifacts required to reproduce our case study for LLM-based log interpretation, including prompts and model responses\footnote{\url{https://github.com/ait-aecid/attack-manifestations-interpretation}}. We summarize our contributions as follows. 

\begin{itemize}
	\item A labeled data set of cyber attack manifestations comprising system logs and security alerts,
	\item a systematic analysis and categorization of the collected attack manifestations, and
	\item an illustrative evaluation of LLM-based log interpretation using the proposed data set.
\end{itemize}

The remainder of this paper is structured as follows. Section \ref{related} summarizes related work on LLM-based interpretation of system logs and security alerts and furthermore reviews existing data sets. Section \ref{generation} describes the methodology for generating the data sets introduced in this paper. Section \ref{analysis} analyzes the captured attack manifestations extracted from the data. Section \ref{interpretation} presents an illustrative evaluation of LLM-based interpretation. Section \ref{discussion} discusses the implications and limitations of our work and outlines directions for future research. Finally, Section \ref{conclusion} concludes the paper.

\section{Related Work} \label{related}

In this section we first summarize relevant publications in the area of system log and alert interpretation using LLMs. Subsequently, we provide a review of public data sets that are useful to evaluate such approaches.

\subsection{LLM-based Interpretation}

We first discuss approaches that apply LLMs to generic system logs and then focus on work conducted in the cyber security domain.

\subsubsection{Analysis of Generic Log Data} \label{genericlog}

The rapid emergence and advancement of LLMs have influenced a wide range of research fields, including system log analysis. Since logs are typically available in unstructured or highly heterogeneous formats, LLMs appear as a natural candidate to assist human analysts in extracting actionable insights from such data. One of the earliest systematic investigations in this area is presented by Mudgal et al. \cite{mudgal2023assessment}, who evaluate the extent to which LLMs can infer information directly from raw log data. They consider the following key analysis tasks: \textit{Log Parsing}, which aims to derive log templates and extract parameters, \textit{Log Analytics}, which involves disciplines such as anomaly detection, root cause analysis, and event prediction, and \textit{Log Summarization}, which provides concise summaries of large amounts of log data. In addition to parsing and anomaly detection, the study of Liu et al. \cite{liu2025loglm} considers \textit{Log Interpretation} as the generation of natural-language explanations of the semantic meaning of log messages, as well as \textit{Root Cause Analysis} and \textit{Solution Recommendation} as post-detection activities that aim to support analysts with  actionable advice for the resolution of potential issues.

Several recent publications present approaches that automate some or all of these log analysis tasks. Liu et al. \cite{liu2024logprompt} propose LogPrompt, a framework for anomaly detection in log data using tailored prompts for ChatGPT. In their subsequent work, the authors propose LogLM \cite{liu2025loglm}, a model designed to support all major log analysis tasks within a unified framework. Shanto et al. \cite{shanto2024console} develop a domain-specific LLM trained explicitly to explain error messages in console logs. Qi et al. \cite{qi2023loggpt} introduce LogGPT, which leverages ChatGPT to detect log anomalies in a zero-shot setting, where no training data is provided, as well as a few-shot setting, where labeled sample data is available to the model. While many studies primarily evaluate detection accuracy, Zhang et al. \cite{zhang2025llm} emphasize the practical importance of providing meaningful interpretations alongside detections. To address this gap, they fine-tune a model termed LLM-LADE, which jointly performs anomaly detection and explanation.

\subsubsection{Analysis of Security Logs and Alerts}

All approaches discussed in the previous section are evaluated on data sets that comprise log data corresponding to normal system or user behavior as well as failure or errors logs. A key difference to log data analysis in the security context is that data sets comprise attack traces that result from deliberate and adversarial actions rather than unintentional failures. While some concepts from anomaly detection and interpretation are applicable in both domains, correct classification and reasoning about attack manifestations generally requires to understand an attacker's intent, place each attack step in the context of an attack chain, and incorporate external knowledge from attack frameworks and tools.

Karlsen et al. \cite{karlsen2024benchmarking} evaluate LLMs on both generic and security-focused data sets and show that fine-tuning substantially improves detection performance across domains. Additionally, they demonstrate how LLMs can highlight those parts of log entries that contribute most strongly to a classification decision, thereby enhancing model transparency for human analysts. Palma et al. \cite{palma2025leveraging} address the high computational and financial costs associated with large proprietary models and instead evaluate a locally deployed language model for log-based detection and interpretation. Sun et al. \cite{sun2025alerts} present a framework for host-based intrusion detection that relies on analysis of raw system logs by LLMs. Their work addresses common issues for such tasks, such as context window limitations that prevent analysis of large chunks of log data.

Beyond log explanation, LLMs have also been applied to map low-level log data to higher-level threat intelligence concepts. Cotti et al. \cite{cotti2025ontologx} present OntoLogX, which constructs ontology-grounded knowledge graphs from raw logs and uses them to predict MITRE ATT\&CK tactics. Tejero et al. \cite{tejero2025evaluating} demonstrate that LLMs outperform conventional machine learning techniques for the task of attack classification in Internet of Things (IoT) logs. Moreover, they map attack classes to entries of the CAPEC\footnote{\url{https://capec.mitre.org/}} attack pattern database and thereby show that LLMs are able to enrich detections with mitigation recommendations tailored to IoT environments.

The interpretive capabilities of LLMs are also suitable to explain intrusion detection alerts. Balogh et al. \cite{balogh2024using} employ LLMs to analyze alerts generated by a network intrusion detection system and map them to MITRE ATT\&CK techniques. Although some correct classifications are achieved, the authors conclude that the models are currently insufficiently reliable and too resource-intensive for automated deployment in practice. In contrast, Hans et al. \cite{hans2025security} report high classification accuracy achieved on similar data sets. In addition to mapping alerts to tactics and techniques, they investigate alert grouping, i.e., segmenting continuous alert streams into clusters representing individual attacker actions or attack steps. Zha et al. \cite{zha2024leveraging} also address the alert grouping problem by using LLMs to infer the root causes of alerts, specifically to trace cascading effects of service failures that often cause alert floods.

\subsection{Data Sets} \label{datasets}

\begin{sidewaystable}[!thbp]
	\centering
	\tiny
	\setlength{\tabcolsep}{4pt}
    \caption{Overview of reviewed data sets.}
    \begin{tabular}{lccccccccccc}
		\toprule
		\textbf{Dataset} & \textbf{Year} & \textbf{Labels} & \textbf{Generation} & \makecell{\textbf{Number of} \\ \textbf{Tactics}} & \makecell{\textbf{Number of} \\ \textbf{Techniques}} & \makecell{\textbf{System} \\ \textbf{Logs}} & \makecell{\textbf{Network} \\ \textbf{Traffic}} & \makecell{\textbf{HIDS} \\ \textbf{Alerts}} & \makecell{\textbf{NIDS} \\ \textbf{Alerts}} & \textbf{Environment} & \textbf{Network} \\
		\midrule
		Supercomputer Logs \cite{oliner2007supercomputers} & 2006 & Failure Type & Real & 0 & 0 & \checkmark & & & & Supercomputer & Generic \\
		CICIDS2017 \cite{sharafaldin2018toward} & 2017 & \makecell{Attack Type} & Scripted & 10* & 19* & \checkmark & \checkmark & & & Windows + Linux & Enterprise \\ 
		Loghub \cite{zhu2023loghub} & 2019 & Various & Mixed & 0 & 0 & \checkmark & & & & Various & Various \\ 
		Mordor/OTRF \cite{mordor} & 2019 & Technique & Scripted & 9 & 35 & \checkmark & \checkmark & & & Windows + Linux & Enterprise \\ 
		AIT-LDSv1.1 \cite{landauer2020have} & 2020 & Attack Type & Scripted & 7* & 15* & \checkmark & & & & Linux & Enterprise \\ 
		CYSAS-S3 \cite{medenou2020cysas} & 2020 & Attack Type & Scripted & 8 & 11 & \checkmark & \checkmark & \checkmark & \checkmark & Windows + Linux & Military \\ 
		DAPT-2020 \cite{myneni2020dapt} & 2020 & \makecell{Attack Type} & Red Team & 4 & 16 & \checkmark & \checkmark & & \checkmark & Linux & Enterprise \\ 
		EVTX-ATTACK-SAMPLES \cite{evtxas} & 2020 & Technique & Scripted & 11 & 58 & \checkmark & & & & Windows & Enterprise \\ 
		PWNJUTSU \cite{berady2022pwnjutsu} & 2022 & None & Red Team & 5 & 13 & \checkmark & \checkmark & & & Windows + Linux & Generic \\ 
		UWF-ZeekData22 \cite{bagui2023introducing} & 2022 & Technique & Red Team & 14 & 36 & & \checkmark & & & Windows + Linux & Generic \\ 
		AIT-LDSv2.0 + AIT-ADS \cite{landauer2022maintainable, landauer2024introducing} & 2022 & Attack Type & Scripted & 8 & 10 & \checkmark & \checkmark & \checkmark & \checkmark & Linux & Enterprise \\
		Unraveled \cite{myneni2023unraveled} & 2023 & Technique & Manual & 5 & 12 & \checkmark & \checkmark & & \checkmark & Windows + Linux & Enterprise \\ 
		PicoDomain \cite{laprade2020picodomain} & 2023 & Attack Type & Scripted & 8* & 19* & & \checkmark & & & Windows & Enterprise \\ 
		LMDG \cite{mabrouk2025lmdg} & 2024 & Attack Type & Scripted & 5* & 10* & \checkmark & \checkmark & & & Windows & Enterprise \\ 
		UWF-ZeekData24 \cite{elam2025introducing} & 2024 & Technique & Scripted & 7 & 5 & & \checkmark & & & Windows + Linux & Enterprise \\ 
		APT-Persistence \cite{rahal2025dataset} & 2024 & None & Scripted & 6* & 11* & \checkmark & & \checkmark & & Windows & Enterprise \\ 
		ATLASv2 \cite{riddle2023atlasv2} & 2024 & Attack Type & Manual & 13* & 30* & \checkmark & & \checkmark & & Windows & Generic \\ 
		EVTX-to-MITRE-Attack \cite{evtxtoma} & 2024 & Technique & Scripted & 11 & 49 & \checkmark & & \checkmark & & Windows & Enterprise \\ 
		Linux-APT \cite{karim2024advanced} & 2024 & Technique & Manual & 12 & 26 & \checkmark & & \checkmark & & Linux & Generic \\ 
		Atomic-EVTX \cite{atomic} & 2025 & Technique & Scripted & 13 & 121 & \checkmark & & & & Windows & Generic \\
		CasinoLimit \cite{kilian2025casinolimit} & 2025 & Technique & Red Team & 14 & 67 & \checkmark & \checkmark & & \checkmark & Linux & Enterprise \\ 
		COMISET \cite{perez2025comiset} & 2025 & Technique & Red Team & 11 & 50 & \checkmark & & & & Windows & Enterprise \\ 
		CIDDS-113 \cite{wolf2026implications} & 2026 & Attack Type & Scripted & 5* & 6* & \checkmark & \checkmark & \checkmark & & Windows + Linux & Enterprise \\ 
		\textbf{CAM-LDS (Our Dataset)} & 2026 & Technique & Scripted & 13 & 81 & \checkmark & \checkmark & \checkmark & \checkmark & Linux & Enterprise \\
		\bottomrule
	\end{tabular}
	\vfill
	\textit{* Asterisks indicate that tactics and technique counts are not explicitly stated in the paper and thus estimated based on attack descriptions.}
	\label{tab:dataset_comparison}
\end{sidewaystable}

Publicly available log data sets play a central role in the development and evaluation of methods for log analytics, as they enable reproducible experimentation and fair comparison across approaches. Historically, however, high-quality data sets have been sparse, particular those that fulfill common requirements such as correctness, relevance, realism, and timeliness \cite{landauer2022maintainable, bonninghausen2024introducing}. In recent years, several log data sets have emerged as de facto benchmarks, and a number of newly published data sets constitute suitable candidates for specific research directions, including LLM-based log interpretation. In the following, we review existing system log data sets, compare their key characteristics, and highlight notable properties of individual contributions. 

Table \ref{tab:dataset_comparison} summarizes relevant properties of all reviewed data sets, including the types of collected data, the approach to attack generation, the granularity of labels, the operating systems from which logs are captured, the emulated network topology, and the number of covered attack tactics and techniques corresponding to the MITRE ATT\&CK framework. For comparability, we only count techniques, because not all data sets provide labels at the sub-technique level; where necessary, sub-techniques are therefore mapped to their corresponding techniques.

Some of the most widely used log data sets, in particular in the research area of anomaly detection, are provided in the LogHub repository \cite{zhu2023loghub} as well as the collection of supercomputer logs originally described by Oliner et al. \cite{oliner2007supercomputers}. These log data sets primarily comprise logs from normal system behavior as well as events that indicate failures; however, there are no traces of cyber attacks present in the data. Consequently, their relevance to cyber-security–focused research is limited. Nonetheless, they are suitable for evaluation of generic approaches that leverage LLMs for anomaly detection or log interpretation as presented in Sect. \ref{genericlog}.

Several data sets have been collected specifically for evaluating intrusion detection systems, with a primary focus on network traffic \cite{goldschmidt2025network}. For example, the authors of PicoDomain \cite{laprade2020picodomain} simulate an enterprise environment and prepare a kill chain comprising client-side exploitation, privilege escalation, and data exfiltration. Data collection relies on the network monitoring tool Zeek\footnote{\url{https://zeek.org/}}, which produces network-centric log files. The CICIDS2017 \cite{sharafaldin2018toward} data set includes both network traffic and system log data; however, since only network traffic is labeled, it is predominantly used for network intrusion detection. A similar limitation applies to DAPT-2020 \cite{myneni2020dapt}, which collects system logs from various sources such as syslog and audit, but only provides labels for network traffic. 

The UWF-ZeekData22 \cite{bagui2023introducing} data set was collected during a cyber security exercise in which students employed both defensive and offensive tools. While the labeled data set comprises attacks from all 14 MITRE ATT\&CK tactics, the vast majority of events corresponds to reconnaissance activities such as port scanning, and data collection is limited to Zeek logs. UWF-ZeekData24 \cite{elam2025introducing} is a follow-up data set to UWF-ZeekData22 that replaces student-driven attacks with scripted executions, which reduces timestamp inaccuracies, labeling errors, and noise according to the authors.

Beyond network-centric data sets, several data sets that include system logs rely on red teaming exercises for attack execution. The PWNJUTSU data set \cite{berady2022pwnjutsu} contains attack traces generated by 22 offensive security experts and includes both network traffic and diverse system logs, such as Linux authentication, access, and audit logs, as well as Windows event logs. The authors of CasinoLimit \cite{kilian2025casinolimit} emphasize that capturing both network traffic and system logs is important to reconstruct attack scenarios. They emulate the network infrastructure of a casino and task 114 penetration testers with attacking the environment. Subsequently, they manually label the resulting data set with 67 distinct attack techniques observed over a time span of 10 hours. The authors of COMISET \cite{perez2025comiset} rely on rule-based labeling of attack traces produced by a red team targeting an emulated small enterprise network of Windows hosts, yielding 50 techniques across 11 tactics in two environments.

While red-teaming and penetration-testing campaigns typically provide broad coverage of attack behaviors \cite{elam2025introducing}, data sets that focus on specific tactics and settings often rely on scripted attacker behavior. For example, the APT-Persistence data set \cite{rahal2025dataset} concentrates on persistence techniques, LMDG \cite{mabrouk2025lmdg} focuses on lateral movement, CIDDS-113 \cite{wolf2026implications} centers on file-encryption attacks, and CYSAS-S3 \cite{medenou2020cysas} targets a military-oriented use case.

Most existing data sets, however, model attack scenarios that follow a full or partial kill chain and therefore span multiple tactics. AIT-LDSv1.1 \cite{landauer2020have} includes network scanning, account discovery, brute-force initial access, exploitation for persistence, and privilege escalation, and was designed to support the evaluation of host-based intrusion detection systems. A distinguishing feature of this data set is the collection of fine-grained Linux audit logs. The same authors introduce AIT-LDSv2.0, which includes additional attack vectors, collects logs from multiple hosts, and improves event labeling. In another follow-up publication, the authors present AIT-ADS \cite{landauer2024introducing}, an alert data set generated by forensically processing the AIT-LDSv2.0 with signature- and anomaly-based intrusion detection systems. The attack chain emulated in the ATLASv2 \cite{riddle2023atlasv2} data set relies on phishing for initial access and aims at the exfiltration of documents. Captured data sources involve audit logs, browser application logs, DNS logs, and security alerts from a host-based intrusion detection system. Notably, the authors emphasize manual execution of the attack steps to enhance realism.

While the aforementioned data sets employ descriptive attack type labels, several other data sets adopt tactic- and technique-level labeling aligned with the the MITRE ATT\&CK framework. This includes several publicly available data sets not directly associated with research publications, such as Atomic-EVTX \cite{atomic}, EVTX-to-MITRE-Attack \cite{evtxtoma}, and EVTX-ATTACK-SAMPLES \cite{evtxas}. While these data sets cover a large number of techniques, they are restricted to Windows environments. The Mordor/OTRF \cite{mordor} data set extends coverage to Linux hosts and web services in addition to Windows event logs. Unraveled \cite{myneni2023unraveled} builds upon DAPT-2020 \cite{myneni2020dapt} by expanding attack tactics, improving IDS deployment, and refining labels. Linux-APT \cite{karim2024advanced} simulates the behavior of selected APT groups and provides technique-level labels of attacks against Linux-based environments.

\begin{figure*}
	\centering
	\includegraphics[width=\textwidth]{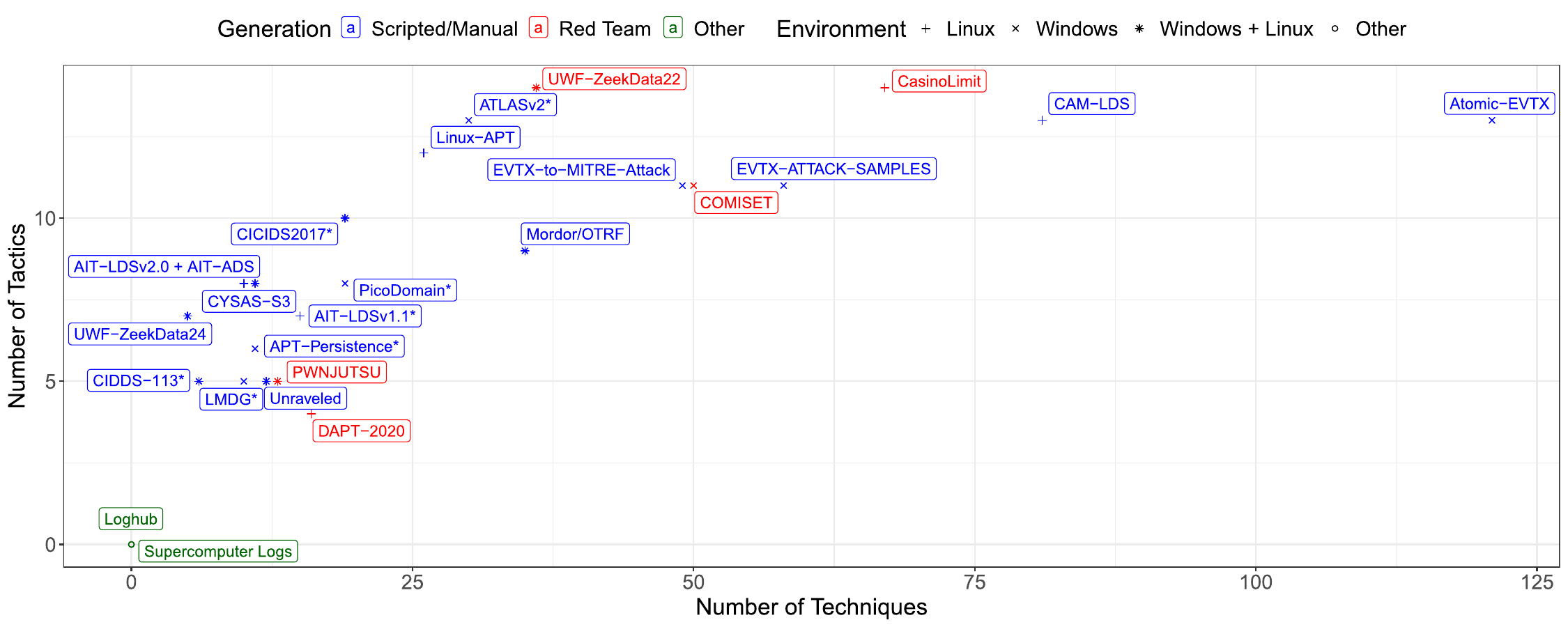}
	\caption{Number of MITRE ATT\&CK tactics and techniques covered by the reviewed data sets. Asterisks indicate that tactics and technique counts are not explicitly stated in the paper and thus estimated based on attack descriptions.}
	\label{fig:related}
\end{figure*}

In summary, several existing data sets are ill-suited for log interpretation in a cyber-security context, either because they do not contain attack traces or because they focus primarily on network traffic rather than system logs. As illustrated in Figure \ref{fig:related}, many data sets cover fewer than nine tactics and fewer than twenty techniques, which constrains the scope of evaluations conducted on them. Data sets with broader coverage of attack techniques typically originate from red teaming exercises, which introducing challenges related to controllability, reproducibility, and labeling reliability \cite{elam2025introducing}, or are limited to Windows-only environments, thereby excluding Linux-specific log sources.

To address these limitations, we introduce CAM-LDS, a system log data set collected through seven fully scripted attack scenarios that cover 81 distinct attack techniques executed in a reproducible Linux environment. Accordingly, we consider it as a complementary data set to existing Windows-centric data collections. In addition to system logs and network traffic, CAM-LDS includes alerts generated by host- and network-based intrusion detection systems, as well as system configuration data that provides valuable contextual information for log interpretation and attack analysis.

\section{Data Set Generation} \label{generation}

This section describes our approach to generate the CAM-LDS. After explaining our methodology, we present the network topology common to all scenarios, the adversarial emulation tool used to execute the attacks, and the corresponding attack chains of each scenario.

\subsection{Methodology} \label{procedure}

\begin{figure}
	\centering
	\includegraphics[width=.49\textwidth]{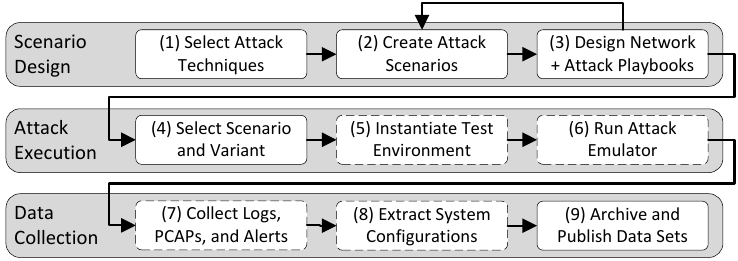}
	\caption{Overview of our methodology. Dashed boxes indicate automated steps that require minimal manual intervention.}
	\label{fig:meth}
\end{figure}

In this section, we describe our methodology for generating log data sets. We thereby refer to the Steps (1)-(9) displayed in Fig. \ref{fig:meth}, which summarizes our methodology.

\subsubsection{Scenario Design}

Our methodology is driven by the objective of collecting log data corresponding to a wide variety of attack techniques. Accordingly, Step (1) of our methodology focuses on the systematic selection of suitable techniques from the MITRE ATT\&CK framework, one of the most widely adopted and up-to-date references for modeling adversary behavior. At the time of writing, the ATT\&CK Enterprise matrix comprises 216 distinct techniques\footnote{This work relies on version 18.1 of the MITRE ATT\&CK Enterprise matrix, which is the most recent version available at the time of writing.}. We review each technique individually and assess it according to the following criteria: (i) \textit{Observability}. We base our selection on the ``Detection Strategies'' specified for each technique in the MITRE ATT\&CK framework. These recommendations describe how adversary behavior can be detected in practice and often indicate relevant data sources. We therefore focus on techniques whose detection strategies suggest the generation or modification of observable artifacts in system or application logs. Techniques that do not produce such traces are of limited relevance for log-centric analysis and are consequently excluded. For example, we exclude most techniques associated with the ``Resource Development'' tactic, which primarily describes activities that occur outside of the target infrastructure, such as acquiring domains or social media accounts. (ii) \textit{Feasibility}. We restrict our selection to techniques that can be realistically emulated within our Linux-based test environment. As a result, techniques that are specific to other platforms, such as the Windows operating system, various mobile systems, or cloud-native environments, are excluded. (iii) \textit{Automation}. We require that execution of the attack technique can be fully automated to support scripted attack chains and ensure reproducibility. Consequently, techniques that inherently involve manual interaction with physical hardware, such as inserting USB devices, are omitted.

After applying these criteria to the entire framework, we end up with a set of 122 attack techniques for potential implementation. Thereby, in order to construct a challenging data set where some individual attack steps are only recognizable as such when considered in the context of the entire kill chain, we purposefully keep techniques that could yield manifestations similar to those of benign system behavior. In particular, this includes activities commonly performed by administrators, such as system information collection.

Step (2) aims to create attack scenarios based on the initial set of 122 attack techniques. We start by grouping techniques according to their infrastructure requirements, such as the need for specific services or particular network configurations. Subsequently, we arrange the grouped techniques into coherent and realistic attack sequences, following common kill chain models and reflecting the complexity of real-world attacks in terms of technique diversity \cite{shen2024decoding}, ultimately resulting in seven attack scenarios. At this stage, we exclude some techniques that cannot be meaningfully integrated into any scenario. We further observe that several techniques are functionally interchangeable, and therefore introduce variants in our scenarios, where individual attack steps are realized using different techniques while achieving the same attacker objective. After completing this step, we obtain a conceptual specification of both the required network infrastructure and the complete set of attack scenarios.

The technical realization of the infrastructure and the implementation of the attack emulation playbooks in Step (3) is carried out as an iterative task that is coupled with Step (2). The reason for this is that during development and testing, techniques may be moved between or duplicated to other scenarios to produce more informative or diverse log manifestations. Moreover, certain techniques that turn out impractical or unsuitable for the environment or attack chains are removed, and the scenarios are revised accordingly. Note that during this phase we also carry out validation activities to ensure that all scenarios are logically consistent, correctly implemented, and accurately labeled with the corresponding attack techniques. Thereby, all attack playbooks and scripts are independently cross-checked by two other experts. After completion of the implementation phase, we eventually end up with a fully scripted infrastructure as well as playbooks for 81 attack techniques across seven scenarios.

\subsubsection{Attack Execution}

In Step (4), we select one of the previously designed attack scenarios, which are accompanied by configuration files that specify the required network topology as well as the necessary software components and versions. Note that the subsequent steps refer to a single simulation run; however, we repeat these steps for every scenario and for all possible combinations of scenario variants. Step (5) instantiates the infrastructure for the selected scenario in a virtualized environment, which is done from scratch for every scenario to avoid that any traces left from previous attacks remain on the systems and to ensure that the infrastructure is in a predictable state. Deployment of the network and all hosts is fully automated, as the entire infrastructure is specified using Infrastructure-as-Code. We employ OpenTofu\footnote{\url{https://opentofu.org/}} and Terragrunt\footnote{\url{https://terragrunt.gruntwork.io/}} to provision the network components and virtual machines, and use the configuration management framework Ansible\footnote{\url{http://www.ansible.com/}} to install and configure all software components on the deployed hosts.

Step (6) executes the attack emulation for the selected scenario and, where applicable, its variants. Note that we do not start the emulation tool immediately after infrastructure deployment, as newly instantiated systems may generate background log events during and shortly after startup. To prevent such events from overlapping with attack traces, we introduce a stabilization period of 15 minutes before starting the attack emulation. In addition, log events may be generated with slight delays following individual attack actions, for instance, due to communication overheads or buffering when large numbers of events are generated. To account for this, we enforce a minimum sleep interval of 15 seconds between consecutive attack steps.

\subsubsection{Data Collection} \label{collection}

After completion of the attack execution, Step (7) collects log data from 18 distinct data sources and across all hosts in the environment. This includes standard Linux log sources such as audit logs, system logs (syslog), authentication logs, access and error logs, package management logs, and CRON logs. In addition, we collect application-specific logs from services deployed in the scenarios, including Zoneminder, FTP, Puppet, Docker, Nextcloud, and the mail server, as well as system performance metrics. We further collect alerts generated by the host-based intrusion detection system Wazuh\footnote{\url{https://wazuh.com/}}, which we utilize with its default rule set. Netflows and network packet captures are captured using Suricata\footnote{\url{https://suricata.io/}}, which simultaneously serves as our network-based intrusion detection system and is also used with standard detection rules.

Step (8) extracts system configurations from all hosts. These configurations provide important contextual information for forensic analysis, such as determining the presence of vulnerable software versions or insecure system settings. In the final Step (9), we archive all collected data and release them in their raw format without modification to ensure that no potentially relevant information is lost. Anonymization is not required, as no human users interact with the infrastructure during the simulations.

\subsection{Network Topology} \label{infrastructure}

\begin{figure*}
	\centering
	\includegraphics[width=.8\textwidth]{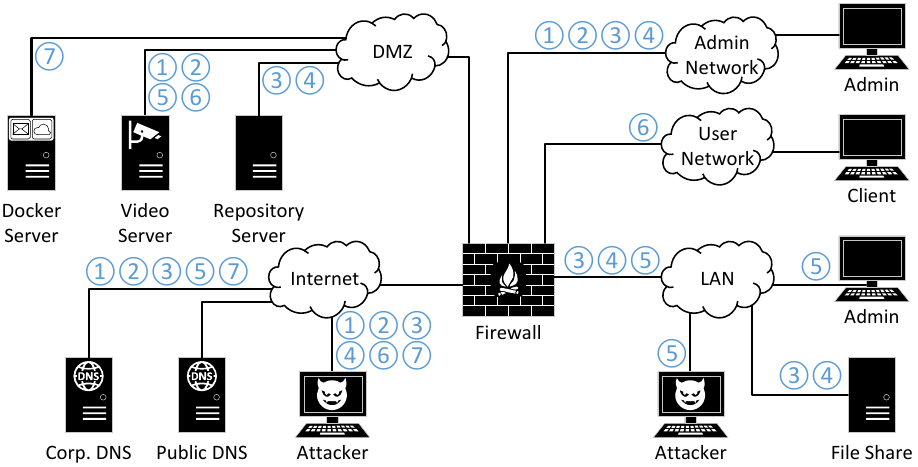}
	\caption{General network topology and scenario-specific deployments of components.}
	\label{fig:network}
\end{figure*}

Cyber attacks rarely affect a single host in isolation; instead, their traces typically manifest across multiple interconnected systems. Gaining a comprehensive understanding of attacker behavior, e.g., as part of forensic investigations conducted after an incident, therefore requires correlating log data from multiple hosts and infrastructure components. To capture attack traces from a realistic setup of such a distributed systems, we design a network topology that approximates the IT infrastructure of a small enterprise as illustrated in Fig. \ref{fig:network}.

Following established best practices for network segmentation \cite{ISO27033}, the topology is divided into five security zones that are interconnected via a firewall: (i) \textit{Internet}. This zone provides external connectivity and contains a public DNS server as well as a corporate DNS server for the simulated organization. In all but one of our scenarios, the attacker gains initial access through the Internet zone. (ii) \textit{Demilitarized Zone (DMZ)}. The DMZ hosts several publicly exposed services, including a Docker server running a mail platform and a cloud service as containers, a repository server used for internal asset management, and a video server providing video surveillance functionality. (iii) \textit{Local Area Network (LAN)}. The LAN provides internal services to the organization, including a file share accessible to employees. (iv) \textit{User Network}. This zone represents the standard working environment of employees, who access the enterprise network using their company workstations. (v) \textit{Admin Network}. The admin network is reserved for network administrators and serves as the default zone for privileged access to the infrastructure.

The network is instantiated on a per-scenario basis, deploying only those components that are required for executing the attack and for collecting the corresponding traces. Figure \ref{fig:network} shows which components are deployed in each scenario by referencing the number of the respective scenario in blue circles. For example, in Scenario 5 we do not deploy the file share, but assume that a network administrator is connected to the LAN. In this scenario, the attacker also accesses the network through the LAN rather than the Internet. The next section describes our method for emulating the attacker's behavior in detail.

\subsection{Attack Emulation} \label{attackmate}

Our attack emulation strategy is guided by two primary requirements: repeatability and realism of the generated log artifacts. Repeatability is essential to enable multiple executions of identical attack scenarios, which in turn allows to verify the consistent generation of attack manifestations across runs. Moreover, repeatability enables automation of attack execution, thereby reducing manual effort during data collection and minimizing inconsistencies as well as human error. Furthermore, automated and repeatable attack scenarios facilitate the systematic generation of variants in which individual steps of an attack chain are replaced or modified, allowing us to cover additional attack techniques in a controlled manner. Since neither red teaming exercises nor manual attack execution (cf. Sect. \ref{datasets}) enable repeatable and automated experiments that yield consistent and reproducible results \cite{elam2025introducing}, we opt for script-based attack execution.

Naive automation through scripts can negatively affect the realism of observed attack traces \cite{riddle2023atlasv2, landauer2026attackmate}. Specifically, it is non-trivial to realize system interactions with simple shell scripts in such a way that they are indistinguishable from actual attacker behavior. For example, human users generally rely on interactive programs such as text editors to modify files, while scripts are limited to non-interactive mechanisms such as file replacement and stream editing to achieve equivalent actions. These differences are visible in system logs and may lead to artifacts that are uncharacteristic of real-world attacks, thereby reducing the authenticity of the resulting data sets and possibly affecting the validity of any evaluation results.

\begin{figure}[t]
	\centering
\begin{lstlisting}[language=yaml,escapechar=§]
vars:
  $MAIL_HOST: mail.attackbed.com
  $DOMAIN: attackbed.com
  $DNS_LIST: /usr/local/share/SecLists/Discovery/DNS/subdomains-top1million-5000.txt

commands:
- type: shell
  cmd: dnsenum -f $DNS_LIST --dnsserver 192.42.0.233 $DOMAIN
  metadata:
    techniques: "T1590.002, T1591"
    description: "Enumerate subdomains of corporate domain-zone"

- type: shell
  cmd: ./smtp-user-enum.pl -M VRFY -U smtp-user.txt -t $MAIL_HOST
  metadata:
    techniques: "T1589.002"
    description: "Enumerate email addresses on SMTP-Host"

- type: shell
  cmd: hydra -l "alice@$DOMAIN" -P smtp-pass.txt -c 5 -s 143 $MAIL_HOST imap  
  metadata:
    techniques: "T1078.002, T1110.001, T1133"
    description: "Brute-force IMAP-password using the already enumerated username"
\end{lstlisting}
\caption{Sample AttackMate playbook for DNS enumeration, user identification, and brute-force login.}
\label{fig:playbook}
\end{figure}

To reconcile the requirements of repeatability and realism, we employ AttackMate \cite{landauer2026attackmate}, an open-source adversarial emulation framework designed specifically for reproducible experiments with realistic attack execution across all phases of the kill chain. AttackMate uses fully scripted playbooks to define attack chains in a deterministic and repeatable manner. Figure \ref{fig:playbook} shows an example playbook consisting of three sequential attack steps. Each step executes a predefined command on the target system, yielding consistent log traces as long as the underlying infrastructure remains unchanged. During execution, AttackMate generates a timestamped output file that associates each command with the corresponding attack labels specified in the metadata fields. These time-based labels enable precise correlation between attack steps and collected log data or network traffic.

At the same time, AttackMate preserves realism by natively supporting interactive sessions and sending individual keystrokes that are indistinguishable from manual typing. Unlike many existing adversarial emulation tools that rely on agents installed on target hosts, AttackMate operates exclusively from a dedicated attacker machine and interacts with the environment solely through remote interfaces such as SSH, thereby accurately modeling an external adversary. In addition to shell command executions, AttackMate further provides dedicated modules for common attack vectors and interfaces with offensive tools widely used by adversaries, such as Metasploit\footnote{\url{https://www.metasploit.com/}}.

\subsection{Scenarios} \label{scenarios}

Based on the infrastructure setup described in Sect. \ref{infrastructure}, we implement seven attack scenarios as AttackMate playbooks. The scenarios and their respective steps have been designed to represent realistic kill chains that cover many distinct techniques from the MITRE ATT\&CK framework. Table \ref{tab:scenarios} provides a concise overview of our seven scenarios, including their title, a description summarizing the overall setup and the main attack vectors, the number of variants, the resulting simulated attack paths, and the number of labeled attack steps. In total, our data set comprises 34 simulations and 243 attack steps across all scenarios.

Subsequent sections describe each scenario in detail and include visualizations of the corresponding attack paths. In these state chart diagrams, each block represents one or more attack steps associated with a specific set of technique labels; individual steps correspond to single attacker actions and are identified by the numbers shown within the block. Note that even though execution of some attack techniques only requires a single step, several techniques are realized through sequences of steps and thus grouped for brevity. Furthermore, we indicate the tactics corresponding to the involved techniques using markers above each block, namely Reconnaissance (RCN), Resource Development (RSD), Initial Access (IA), Execution (EXE), Persistence (PST), Privilege Escalation (PVE), Defense Evasion (DEF), Credential Access (CRD), Discovery (DSC), Lateral Movement (LAT), Collection (COL), Command and Control (CNC), Exfiltration (EXF), and Impact (IMP).

\begin{table*}[t]
	\centering
	\tiny
	\caption{Overview of attack scenarios.}
	\begin{tabular}{llp{7cm}lll}
		\toprule
		\textbf{ID} &  \textbf{Name} & \textbf{Description} & \textbf{\#Variants} & \textbf{\#Sim.} & \textbf{\#Steps} \\
		\midrule
		1 & \makecell[tl]{Video Server \\ Exploit} & In addition to some reconnaissance and scanning attacks, this scenario revolves around compromising the video server. A key element in this attack chain is the exploitation of the vulnerable video server software. Subsequently, the attack graph involves five variants that allow the attacker to escalate to root privileges as well as three variants that provide the attacker with persistent access; we simulate all combinations of these variants. & \makecell[tl]{6 (PVE) \\ 3 (PST)} & 18 & 56 \\
		2 & \makecell[tl]{Linux \\ Malware} & The malware considered in this scenario are an implant that establishes reverse-shell connection as well as a rootkit that targets dynamic linking. The remainder of the scenario broadly covers discovery techniques and exfiltrates some data from the compromised machine. & 2 (IA) & 2 & 29 \\
		3 & \makecell[tl]{Lateral \\ Movement} & After logging into the firewall through brute-force password guessing, the attacker moves laterally to the file share in the LAN segment. This is achieved by injecting malicious code either into a shared service file, a software management tool, or a manipulated package. The scenario ends with several destructive activities on the target host, including data encryption and deletion. The entire scenario is implemented for both SSH and VNC connections. & \makecell[tl]{2 (IA) \\ 3 (LAT)} & 6 & 60 \\
		4 & \makecell[tl]{Network \\ Attack} & This scenario aims to compromise the firewall of the network. A port knocking sequence is used to download and run implants, which in turn enable the attacker to modify firewall rules and reach even more hosts in the network. & - & 1 & 22 \\
		5 & \makecell[tl]{Network \\ Sniffing} & In this short scenario, the attacker captures authentication tokens from network traffic and re-uses them to gain access to a system.  & - & 1 & 3 \\
		6 & \makecell[tl]{Attack on \\ Client} & This scenario considers attack cases where a human user is deceived into compromising their own computer, either by downloading and opening a malicious document, installing a fake browser plugin, or providing the attacker access via screen sharing and remote control software. Once inside the system, the attacker installs a backdoor for permanent access. The scenario also involves several activities for data collection, including a keylogger, a data exfiltration tool, and extraction of clipboard data. & \makecell[tl]{3 (IA) \\ 2 (PST)} & 5 & 60 \\
		7 & \makecell[tl]{Docker \\ Attack} & In this scenario, the attacker first carries out some scanning activities and then brute-forces a password of a mail service. By re-using the collected credentials on a cloud storage service, the attacker is able to execute an authenticated exploit that grants them access to the containerized environment that hosts both the mail and cloud storage services. Once inside, the attacker exploits weak configurations to escape the container and obtain root access on the host. & - & 1 & 13 \\
		\bottomrule
	\end{tabular}
	\vfill
	\label{tab:scenarios}
\end{table*}

\subsubsection{Scenario 1: Video Server Exploit} \label{scenario1}

\begin{figure*}
	\centering
	\includegraphics[width=\textwidth]{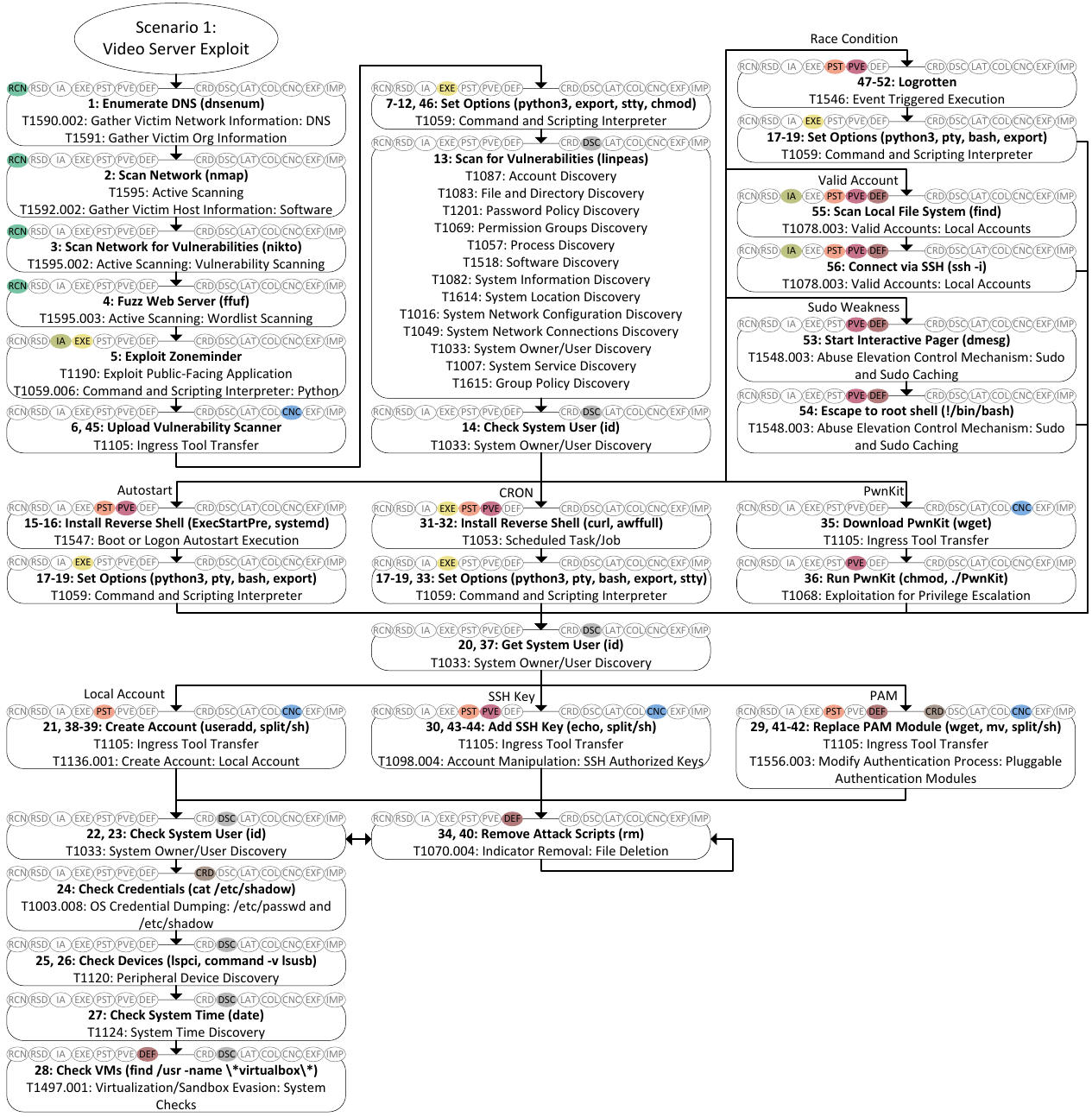}
	\caption{Attack paths of Scenario 1: Video Server Exploit.}
	\label{fig:sc1}
\end{figure*}

As depicted in Fig. \ref{fig:sc1}, Scenario 1 starts with reconnaissance, which is the earliest stage of the kill chain. We do not assume that the attacker has any prior knowledge about the target network beyond its public domain \textit{attackbed.com}. The attacker's first objective is information gathering, which we subdivide into several steps. In the first step, the attacker enumerates subdomains of the corporate DNS server using \textit{dnsenum}\footnote{\url{https://github.com/fwaeytens/dnsenum}}, which reveals \textit{video.attackbed.com}, \textit{fw.attackbed.com}, and \textit{repo.attackbed.com}. The attacker then carries out network mapping of the video subdomain with \textit{nmap}\footnote{\url{https://nmap.org/}}, which is useful to discover available services on hosts, and identifies that port 80 (HTTP) is open on the video server. Next, a network vulnerability scan with \textit{nikto}\footnote{\url{https://cirt.net/nikto/}} shows that the server runs Apache and that the main application entry point is \textit{video.attackbed.com/zm/}, which suggests the presence of the open-source video surveillance software ZoneMinder\footnote{\url{https://zoneminder.com/}}. The attacker then performs deeper web content discovery through fuzzing with \textit{ffuz}\footnote{\url{https://github.com/ffuf/ffuf}}, a fast web fuzzer that enumerates paths using a wordlist.

Since fuzzing does not yield any direct ways to compromise the system, the attacker searches public databases for recent ZoneMinder vulnerabilities and identifies CVE-2023-26035\footnote{\url{https://nvd.nist.gov/vuln/detail/CVE-2023-26035}}, which describes a vulnerability in ZoneMinder’s snapshot action that enables unauthenticated remote code execution. The attacker then resorts to the Metasploit framework, where they find and execute an already implemented exploit for that vulnerability.

With the ability to execute arbitrary commands on the video server, the attacker configures an interactive shell and pursues their new objective of escalating their privileges. To enable fast enumeration of potential privilege escalation vectors, the attacker uploads and executes \textit{linpeas}\footnote{\url{https://github.com/peass-ng/PEASS-ng/tree/master/linPEAS}}, a vulnerability scanner for the Linux operating system that reports misconfigurations, weak permissions, and other common escalation paths as indicated by the broad coverage of attack techniques in Fig. \ref{fig:sc1}. After executing the \textit{id} command to record current user context, the attacker proceeds to escalate their privileges through one of five possible variants modeled in Scenario 1, some of them based on the linpeas output. (i) \textit{Autostart}. The attacker uploads a reverse-shell, opens the ZoneMinder systemd unit file with an interactive text editor, and adds a line that executes the payload on service start, e.g., when the system is booted. When the unit is started, the payload is executed and the attacker gains root access through the reverse-shell. (ii) \textit{CRON}. In this variant the attacker injects commands that download and run a reverse-shell into a system utility that is executed periodically by CRON. For the purpose of Scenario 1, we assume that the \textit{awffull}\footnote{\url{https://packages.debian.org/sid/web/awffull}} tool, a log analysis script that runs under a privileged account and is managed by CRON, was misconfigured by system operators during installation. In particular, vulnerable permissions of the corresponding CRON-job allow unprivileged users to inject commands that are subsequently executed with root permissions. Accordingly, as soon as CRON invokes the manipulated script, the injected commands establish the reverse-shell and provide the attacker with root access to the system. (iii) \textit{PwnKit}. The attacker downloads and runs the \textit{PwnKit}\footnote{\url{https://github.com/ly4k/PwnKit}} script, which is a self-contained exploit for \textit{polkit's pkexec utility}\footnote{\url{https://nvd.nist.gov/vuln/detail/cve-2021-4034}} that provides root access. (iv) \textit{Race Condition}. Most Linux distributions come by default with \textit{logrotate}\footnote{\url{https://linux.die.net/man/8/logrotate}}, a service that enables automatic rotation, compression, and retention of log files. Unfortunately, this tool is vulnerable to a race condition when it runs under root privileges but the managed log files are located in directories accessible to unprivileged users. In this variant, the attacker makes use of the \textit{logrotten}\footnote{\url{https://github.com/whotwagner/logrotten}} exploit, which replaces one of the log files with a symbolic link to an attacker-controlled directory just at the right time so that logrotate creates the new log file in the linked directory. Since the attacker has write access to the file managed by a root process, they can use it to inject a reverse-shell. (v) \textit{Sudo Weakness}. This variant assumes that due to a system misconfiguration, non-privileged users are able to execute \textit{dmesg}, which is a utility to print the kernel ring buffer. By running \textit{dmesg} interactively and abusing its shell-escape features, the attacker is able to spawn a root shell. (vi) \textit{Valid Account}. In this simple variant, the attacker scans the local file system and thereby recovers a private SSH key belonging to a high-privileged user account, which they use to establish a privileged SSH session.

After completing each variant for privilege escalation, the attacker runs the \textit{id} command to confirm their newly obtained privileges. At this stage of the attack chain, the attacker's objective is to obtain persistence to retain access to the system even if the previously exploited vulnerabilities are patched, e.g., when system operators upgrade ZoneMinder to a more recent version that is not vulnerable to the exploit. Scenario 1 involves three variants for persistence that we describe in the following. (i) \textit{Local Account}. The attacker creates a new local user with the name \textit{webadmin} to not trigger any suspicions. Then, the attacker transfers their own SSH key to the authorized keys. (ii) \textit{SSH Key}. This variant is similar to the previous one, with the exception that the attacker copies their own SSH key into the authorized keys of an already existing privileged user. (iii) \textit{PAM}. The attacker replaces the system's \textit{Pluggable Authentication Module} (PAM) with a modified variant that provides backdoor access. 

Note that for some of these variants, the attacker makes use of the split command to obfuscate their activities and reduce traceability. However, since execution of the split command depends on the type of root shell, we are only able to model this optional technique as part of the \textit{PwnKit}, \textit{Sudo Weakness}, and \textit{Valid Account} variants for privilege escalation. All other variants execute the aforementioned commands directly.

After establishing persistence, the attacker runs the id command once more and removes some of the files downloaded as part of the previous attack steps. The attacker then performs targeted post-compromise discovery, in particular, extraction of credentials and hashed passwords for offline cracking, enumeration of connected peripheral devices, checking of system time and timezone, and searching for virtual machines or hypervisor artifacts.

\subsubsection{Scenario 2: Linux Malware} \label{scenario2}

\begin{figure}
	\centering
	\includegraphics[width=.49\textwidth]{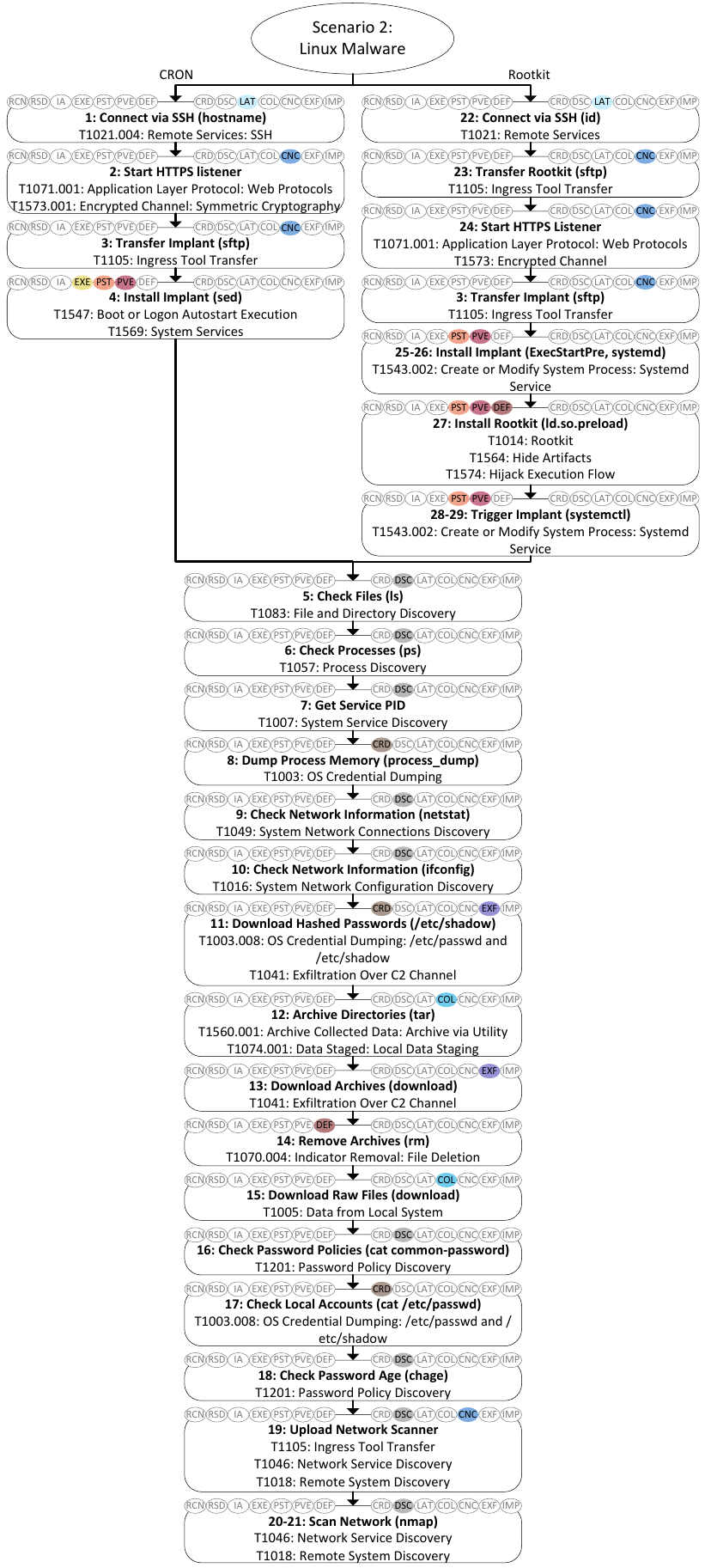}
	\caption{Attack paths of Scenario 2: Linux Malware.}
	\label{fig:sc2}
\end{figure}

Figure \ref{fig:sc2} visualizes the attack chain for Scenario 2. We assume that the attacker has already compromised the video server and obtained root privileges, e.g., through an attack chain comparable to Scenario 1. The attacker's immediate objective in Scenario 2 is to compromise the system further through the installation of malware, which we model in two variants. (i) \textit{CRON}. The attacker connects to the target machine via SSH, starts a HTTPS listener, and transfers an implant via SFTP. The implant is installed by modifying the \textit{awffull} script that has already been targeted in the \textit{CRON} variant of Scenario 1; however, instead of editing the file with a text editor, the attacker uses a stream editor to inject a command that starts the implant. (ii) \textit{Rootkit}. The attacker again connects via SSH, prepares a rootkit, and transfers it via SFTP. Analogous to the previous variant, the attacker also transfers an implant and starts a HTTPS listener. The implant is installed by editing ZoneMinder's systemd unit file similar to the \textit{Autostart} variant of Scenario 1. The rootkit is installed through a dynamic-linker preload technique so that it is loaded system-wide into dynamically linked processes \cite{stuhn2024hidden}. Finally, the attacker reloads systemd and restarts ZoneMinder's service to trigger execution of the implant through systemd.

After completion of either variant, the attacker’s subsequent objective is to collect and exfiltrate relevant information. The post-exploitation phase begins with discovery operations: enumeration of files, running processes, and active network connections, as well as collection of network interface configuration. The attacker then acquires artifacts for offline analysis, including process memory snapshots and a process identifier for the database service. Subsequently, the attacker extracts stored authentication material such as hashed credentials. Collected data is aggregated into archives and extracted; the attacker also removes local artifacts that would reveal the exfiltration event. In addition to bulk collection, the attacker selectively downloads individual files and directories of interest. To support further lateral discovery, the attacker deploys \textit{nmap} that was already used in Scenario 1 and runs it from the compromised host to enumerate reachable hosts and services.

\subsubsection{Scenario 3: Lateral Movement} \label{scenario3}

\begin{figure*}
	\centering
	\includegraphics[width=\textwidth]{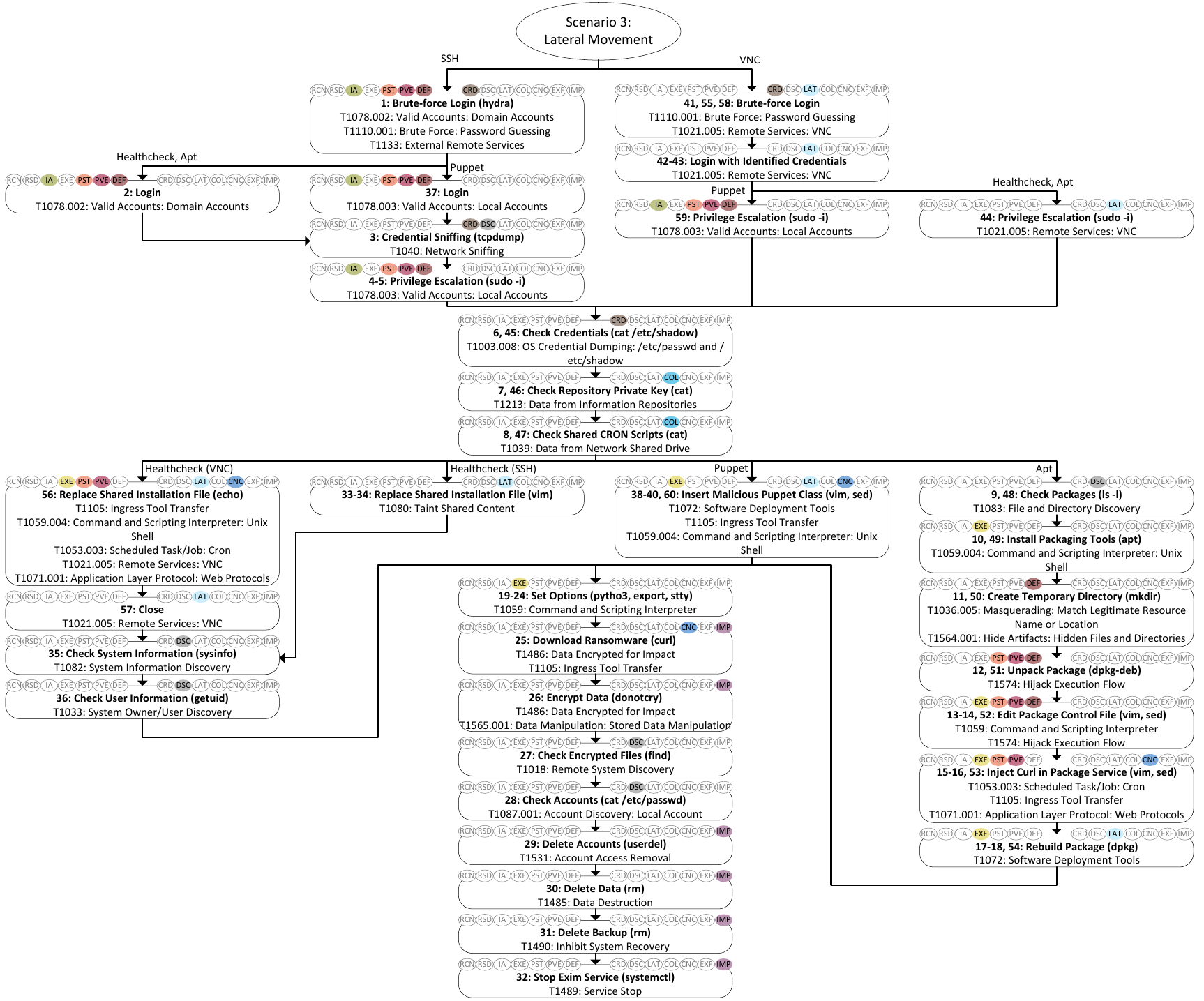}
	\caption{Attack paths of Scenario 3 Lateral Movement.}
	\label{fig:sc3}
\end{figure*}

Scenario 3 emulates how an attacker moves within the network from one host to another, in particular from the initially compromised puppet server to the file share within the LAN segment. Figure \ref{fig:sc3} visually summarizes the entire scenario. The scenario involves two variants regarding the way how the attacker connects to the targeted network. (i) \textit{SSH}. The attacker attempts to log into the puppet server via SSH by brute-forcing several combinations of user names and common passwords. To this end, the attacker employs \textit{hydra}\footnote{\url{https://github.com/vanhauser-thc/thc-hydra}}, a tool that automatically and rapidly goes through a pre-defined file of login credentials and tests whether some of them provide access. For the purpose of our scenario, we assume that one of the users with root privileges has a weak password contained in the list. After logging into the Puppet server, the attacker deploys the network sniffer \textit{tcpdump} to capture credentials transmitted over the network. In our scenario, a scheduled CRON job on the file share periodically retrieves data from the Puppet server via FTP. The attacker intercepts the credentials contained in one of these connections and subsequently re-uses them to escalate to root privileges. (ii) \textit{VNC}. In this variant, the attacker attempts to brute-force a connection via Virtual Network Computing (VNC) rather than SSH. Again, we assume that the attacker manages to guess a weak password of a privileged user after a few attempts, allowing them to log into the puppet server.

Both of the aforementioned variants continue with the attacker printing hashed passwords for offline cracking. Subsequently, they check for the presence of private keys and find one corresponding to the puppet server. Then, the attacker locates a CRON job called \textit{healthcheck} in a Network File System (NSF) share. At this point, we consider several variants how the attacker exploits shared system content to move laterally to another server in the network. Note that while conceptually identical, the technical implementations of these steps are realized differently depending on whether the attacker accesses the system through SSH or VNC as determined by the first variants of the scenario. (i) \textit{Healthcheck}. The attacker edits the \textit{healthcheck} CRON job and injects a line that downloads and establishes a reverse-shell. Once this script is automatically executed on the file share as part of normal \textit{healthcheck} activity, the attacker gains access on that system and has thus achieved lateral movement. (ii) \textit{Puppet}. The attacker realizes that Puppet\footnote{\url{https://github.com/puppetlabs/puppet}}, a service that automates administrative tasks such as server configuration in networks, is installed on the firewall and thus likely used across multiple nodes in the network. The attacker then edits parts of puppet's code; in particular, they inject a malicious class that downloads and executes a reverse-shell and assign that class to the node corresponding to the file share. Next time the Puppet service running on the file share executes that code, the reverse-shell triggers and provides the attacker with access to that system. (iii) \textit{Apt}. The attacker realizes that the puppet server provides APT packages to other servers, including the file share that updates the \textit{healthcheck} package through puppet. They then decide to build a malicious version of that package that appears as an update. This is achieved by downloading packaging tools, unpacking the \textit{healthcheck} package, adding a malicious CRON job that establishes a reverse-shell to the service, and rebuilding the package. As soon as the admin user triggers the installation of the package, e.g., as part of normal updating procedures, the injected code establishes a reverse-shell and thus allows the attacker to move laterally to the file share. Note that we also emulate administrator activities analogous to attacker behavior to ensure that these attack steps succeed.

After completing any of the aforementioned variants, the attacker first configures the shell for interactive access and then continues with a sequence of destructive attack techniques. First, the attacker downloads a ransomware from their own system and uses it to encrypt the media directory on the system. After checking that the files are indeed encrypted, the attacker enumerates all user accounts available on the system and deletes the account of the user whose password is brute-forced at the beginning of this scenario. The attacker proceeds with deleting some directories as well as the system's backup files. Finally, the attacker stops a system service related to mail exchange.

\subsubsection{Scenario 4: Network Attack} \label{scenario4}

\begin{figure}
	\centering
	\includegraphics[width=.4\textwidth]{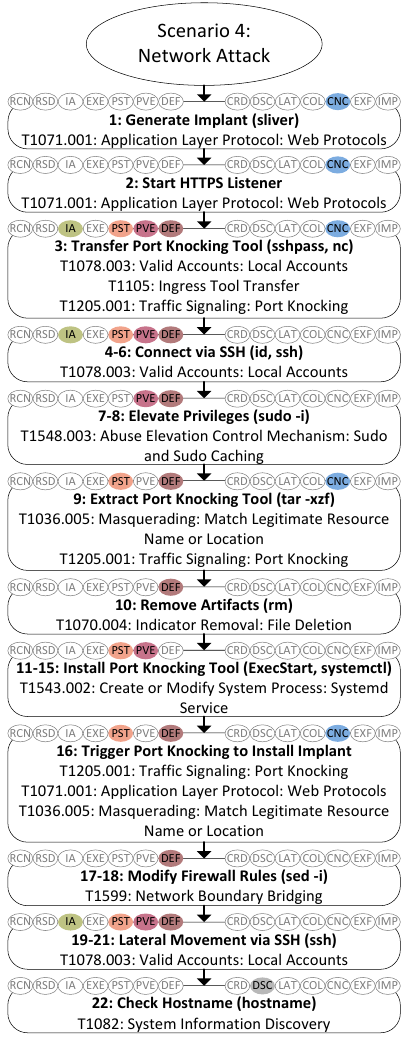}
	\caption{Attack paths of Scenario 4: Network Attack.}
	\label{fig:sc4}
\end{figure}

In Scenario 4, we assume that the attacker has already compromised the repository server within the DMZ and seeks to pivot into and compromise the entire network through targeted attacks against the firewall. The attack chain depicted in Fig. \ref{fig:sc4} starts with the preparation of an implant and an HTTPS listener, similar to the initial steps of Scenario 2. The attacker then re-uses the credentials of a compromised account on the repository server to first transfer a port knocking script to the firewall and then log in as a privileged user. The port knocking script is extracted and installed under the inconspicuous name \textit{auditf} into the location where system binaries are stored; the archive is removed after extraction to reduce forensic traces. The attacker installs the port knocking tool by creating a service that triggers the script on start. Subsequently, they start the service before exiting the firewall and returning to the repository server.

From there, the attacker initiates the port knocking sequence, which causes the previously deployed port knocking tool on the firewall to download and run the implant prepared in one of the earlier steps. Through that implant, which provides remote command execution, the attacker then modifies and reloads the firewall rules so that hosts located in the DMZ are able to connect to hosts in the LAN, thereby effectively enabling to access the fileshare from the repository server. The attacker immediately leverages this ability and connects to the fileshare via SSH, where they print some system information to confirm that they reached the intended target. 

\subsubsection{Scenario 5: Network Sniffing} \label{scenario5}

\begin{figure}
	\centering
	\includegraphics[width=.4\textwidth]{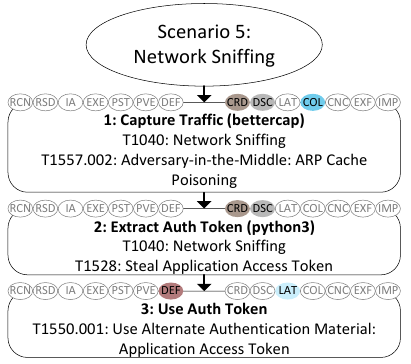}
	\caption{Attack path of Scenario 5: Network Sniffing.}
	\label{fig:sc5}
\end{figure}

Figure \ref{fig:sc5} visualizes Scenario 5, which focuses on network sniffing. The attacker is assumed to have previously deployed a covert network access device in the LAN network, which provides remote access. Implanting this device has been carried out through physical access to one of the hosts within the network, e.g., through an insider who connects such a device to the USB or Ethernet port of an employee's desktop computer; since our focus lies on cyber attacks, this step is considered out of scope of the simulation of Scenario 5. 

Through this device, the attacker launches the sniffing tool \textit{bettercap} for ARP spoofing and to capture network traffic. After some time, we assume that a system administrator connected via LAN logs into the video server located within the DMZ. From the collected data, the attacker is able to extract the administrator's authentication token, which they subsequently re-use to make API calls on the video server.

\subsubsection{Scenario 6: Attack on Client} \label{scenario6}

\begin{figure*}
	\centering
	\includegraphics[width=\textwidth]{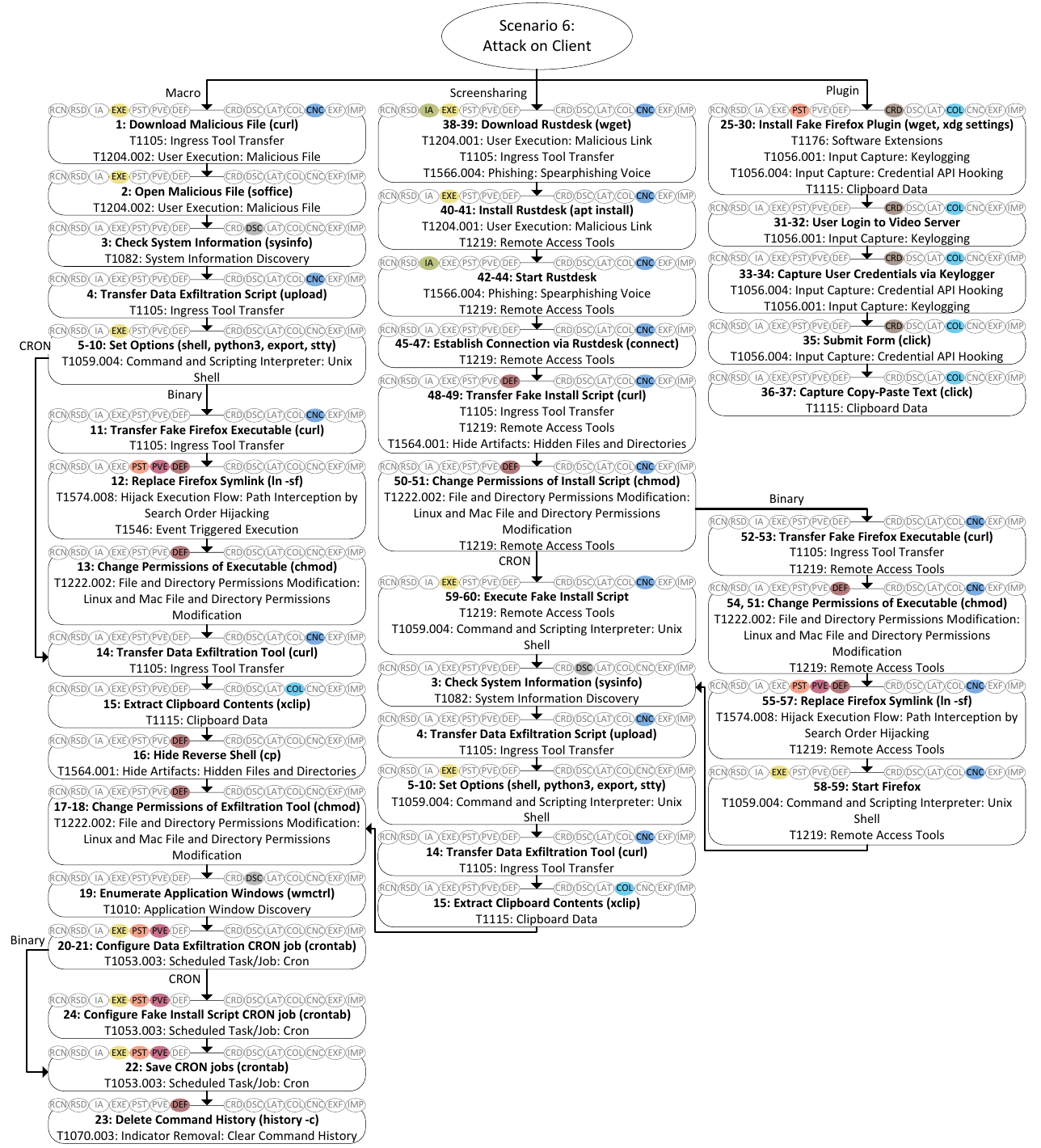}
	\caption{Attack path of Scenario 6: Attack on Client.}
	\label{fig:sc6}
\end{figure*}

Scenario 6 covers three main attack chains targeting a client computer as visualized in Fig. \ref{fig:sc6}. In contrast to previous scenarios, where the attacker primarily relies on tools or exploits of technical vulnerabilities, activities of the human user working on the targeted client computer play a key role for the initial intrusion. Thereby, we assume that the user carries out these activities purposefully but without realizing their malicious nature, for example, through social engineering, where the attacker calls the victim by telephone, pretends to be a technician from the organization's helpdesk, and persuades the user to follow their instructions. Note that in order to avoid that any artifacts from user simulation end up in the log data, we assume that the user remotely accesses their client computer through Virtual Network Computing (VNC). We argue that such remote access software is common in many organizations and does not affect the realism or validity of this attack chain.

Our scenario involves multiple variants; we start our description with the shortest attack chain. (i) \textit{Plugin.} The attacker contacts the user and informs them about the necessary installation of a plugin for the \textit{Firefox} browser as part of a company policy. For the purpose of this scenario, we assume that the user is a web engineer who uses the browser in developer mode, which is necessary for installation. Unknown to them, the plugin is in fact a keylogger crafted by the attacker, who hosts it on their web server under the unsuspecting domain \textit{dailynews-wire.com}. The attacker talks the user through the process of downloading and installing the malicious plugin as well as configuring and restarting the browser. After ending the call, the attacker only needs to wait for the user to type their username and password when logging into any website. As part of our scenario, we assume that the user navigates to the ZoneMinder website and enter their credentials in the login form, which is captured by the keylogger and transmitted to the attacker. As a final step of this attack chain, the user copies and pastes the contents of the browser's search bar; such clipboard data is also captured and transmitted by the plugin.

The other attack chain in this scenario starts with two variants for initial access. (ii) \textit{Macro}. The attacker informs the victim user about a document containing information on organizational policies. Subsequently, the attacker instructs the victim user to open a terminal and download that document from \textit{dailynews-wire.com}, which appears as a valid domain but is in fact hosted by the attacker. Likewise, the downloaded file seems to be a legitimate text document to the unsuspecting user based on its name and contents. However, the attacker previously implanted a malicious macro into the file that is triggered when opening it with the text editing software \textit{LibreOffice}\footnote{\url{https://www.libreoffice.org/}}. When activated, this macro establishes a reverse-shell that provides the attacker with remote access to the system. (iii) \textit{Screensharing}. In this variant, the attacker convinces the user to download and install the legitimate tool \textit{rustdesk}\footnote{\url{https://rustdesk.com/}} under the disguise of helpdesk software. This tool enables to share and remotely control screens, which supposedly allows the attacker to fix some technical issues on the client's computer. As part of this procedure, the attacker instructs the user on how to configure the tool, shares necessary IP addresses and passwords, and tells them to accept the incoming connection when the attacker activates the same tool on their side. Even though this provides the attacker with remote access on the target system, the remote desktop software makes all actions visible on the user's screen, which means that the attacker's capabilities are limited to non-suspicious activities. In addition, their access is only temporary, because the user needs to manually confirm another connection after closing this session. In order to obtain persistent access, the attacker downloads a reverse-shell from their own typosquatted domain \textit{facebock.com}, stores it as a hidden file, and changes its permissions. We assume that these activities do not trigger any suspicions, as the user fails to notice the misspelled domain and expects the alleged technician to diagnose and resolve issues on their computer. 

For the remaining attack chains of \textit{Macro} and \textit{Screensharing} variants we employ similar attack techniques, but slightly vary their order. For brevity, we only explain each step once and refer to Fig. \ref{fig:sc6} for their position in the attack chain of the respective variant.

Both the reverse-shell established through the malicious macro as well as the remote desktop software only provide temporary access that only lasts until the system is rebooted or the session is terminated. We consider two variants how the attacker obtains persistence on the target system. (i) \textit{Browser Binary}. The attacker downloads a binary for a fake executable for the Firefox browser from their own domain, which opens a reverse-shell before starting the browser as usual. Then, they redirect the symbolic link (symlink) of the actual Firefox browser to point to the fake executable. The attacker thus obtains remote access every time when the user opens the browser, even after the system is restarted. (ii) \textit{CRON}. The attacker creates a CRON job that invokes a previously downloaded script that in turn establishes a remote-shell. Note that this CRON job is configured and started at the end of the attack chain alongside other CRON jobs. From that point on, the script is executed every 5 minutes even after the system reboots.

Independent of the aforementioned variant, the attacker runs the \textit{sysinfo} command to validate that the remote-shell works as expected and to obtain some information about the system. The attacker's next objective is to extract data from the compromised system. To this end, they transfer and execute \textit{VeilTransfer}\footnote{\url{https://github.com/infosecn1nja/VeilTransfer}}, a tool that provides several exfiltration techniques and is commonly used in security testing. The attacker uses a script to archive and transfer directories containing potentially sensitive data, in particular, the mail inbox of the Thunderbird application as well as session restore files, key database, and saved credentials of the Firefox browser. At the end of the attack chain, the script is configured as a CRON job that is executed every 10 minutes.

As an additional exfiltration activity, the attacker extracts clipboard contents and enumerates open application windows as part of collection and discovery activities. As a last step of the attack chain, the attacker clears the command history, which makes it more difficult to trace their activities in case that forensic analyses are carried out once the user becomes aware of the attack.

\subsubsection{Scenario 7: Docker Attack} \label{scenario7}

\begin{figure}
	\centering
	\includegraphics[width=.4\textwidth]{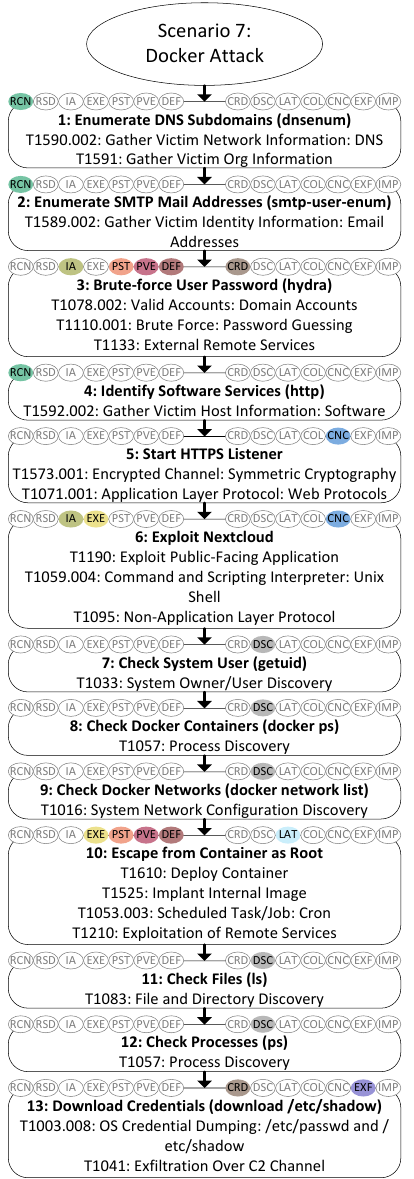}
	\caption{Attack path of Scenario 7: Docker Attack.}
	\label{fig:sc7}
\end{figure}

As visible in the attack chain depicted in Fig. \ref{fig:sc7}, Scenario 7 starts with an enumeration of the corporate DNS servers identical to the first step of Scenario 1 (cf. Sect. \ref{scenario1}). In this case, however, the scan reveals a mail service that is present in the network. The attacker decides to enumerate user names stored on the server using the \textit{smtp-user-enum}\footnote{\url{https://github.com/cytopia/smtp-user-enum}} tool, which yields several results. The attacker chooses user Alice at random and proceeds to brute-force the password of the user's IMAP-account with \textit{hydra}\footnote{\url{https://github.com/vanhauser-thc/thc-hydra}}. Note that in Scenario 3, the same tool is used to brute-force an SSH login (cf. Sect. \ref{scenario3}). We also point out that the AttackMate playbook depicted in Fig. \ref{fig:playbook} illustrates how these initial three steps of the attack chain are implemented for automated execution.

After successfully retrieving the password, the attacker sends a simple web probe to the server and realizes that this host is providing the open-source file sharing platform \textit{Nextcloud}\footnote{\url{https://github.com/nextcloud}} in addition to the mail service. In an attempt to gain access to that system, the attacker executes a Metasploit module that exploits a vulnerable \textit{Nextcloud} application that allows authenticated users to add workflows for command execution\footnote{\url{https://www.rapid7.com/db/modules/exploit/unix/webapp/nextcloud_workflows_rce/}}. In our scenario, this vulnerable application is present on the platform, which means that the attacker is able to successfully execute the exploit using Alice's mail address and password for authentication. 

Through the newly gained shell access, the attacker checks system user information to confirm that they are a non-privileged system user within a containerized environment. At this point, the attacker's primary objective is to escape from the container and gain access on the host running the containers. For the purpose of this scenario, we assume that the Docker API is exposed and accessible from within the container due to weak configurations. This allows the attacker to gather information from the running Docker containers and to enumerate Docker networks. Given this information, the attacker is able to escape the container by temporarily deploying a privileged container through the Docker API, mounting the \textit{etc} directory into that new container, and placing a CRON job that downloads and installs an implant for reverse-shell access into the mounted directory. Once the CRON job triggers, the attacker utilizes the activated implant to access the host system, in particular, to enumerate files and processes as well as to download credential information.

\section{Data Set Analysis} \label{analysis}

This section provides an analysis of the collected log data sets. We first present an overview of the data and then examine how individual attack steps manifest in the collected logs.

\subsection{Data Preparation}

This section describes the data pre-processing steps applied to isolate attack manifestations. We also provide an overview of these manifestations by visualizing log events as timelines and summarizing the covered attack techniques.

\subsubsection{Log Event Filtering} \label{filtering}

Analysis of attack manifestations benefits from isolating log events that are direct consequences of attacker activity. As described in Sect. \ref{procedure}, we aim to achieve such isolation by operating on an idle infrastructure without user interaction and by waiting 15 minutes after system boot before initiating attack execution to avoid interference from startup-related processes. Despite these precautions, an initial inspection of the collected data revealed that even idle systems continuously generate a substantial number of log events that inevitably overlap with attack execution. These background events primarily stem from processes that operate independently of user activity, including periodically scheduled system jobs, time synchronization mechanisms, and sporadically occurring system errors such as timeouts.

To mitigate this issue, we apply a filtering approach that removes log events that correspond to normal system behavior with high probability. Specifically, we consider the ten-minute interval preceding the start of AttackMate execution as a baseline period that captures log activity of an idle system. We compile log events occurring during this period across all hosts and remove timestamps, counters, and random values from these events using regular expressions. We then filter out all log events from the attack execution phase that are identical to entries in this reference set. We acknowledge that some attack steps may also trigger log events that are identical to those observed during normal system operation and may therefore unintentionally be removed by this process. While the informative value of the contextual occurrence of such events could be useful, we argue that their value for analysis is limited compared to those events that remain due to their syntax or parameters.

Even though this automated filtering approach effectively removes a substantial part of normal background events, we still observe certain events that contain highly variable parameters and are thus difficult to normalize with regular expressions. This particularly affects audit logs, for which we thus apply additional manual filtering based on process identifiers, e.g., to remove events related to intrusion detection activity. We refer to our open-source code repository for more information on event filtering. We emphasize that the analyses presented in the following sections only consider the filtered data sets; however, we provide both the filtered and unfiltered data sets in our public repository, allowing researchers to select the version that best fits their use case.

\subsection{Timelines}

\begin{figure*}[!tbhp]
	\centering
	\begin{subfigure}[b]{\textwidth}
		\centering
		\includegraphics[width=\textwidth]{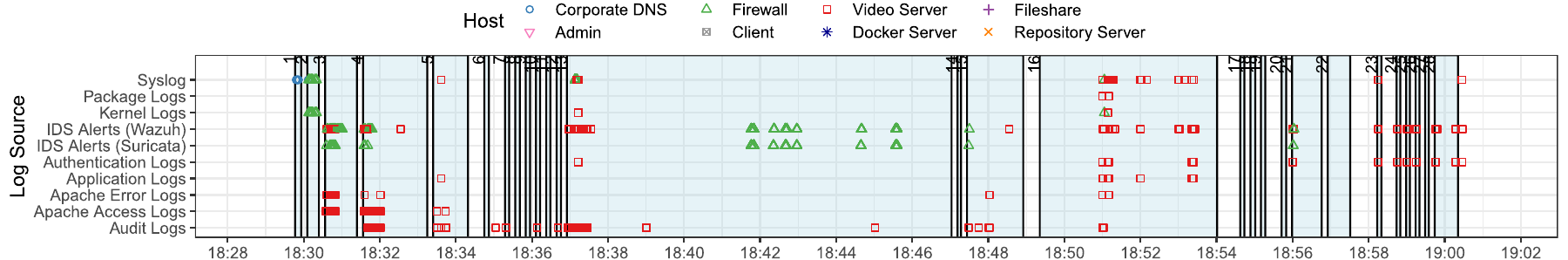}
		\caption{Scenario 1: Video Server Exploit with the Autostart and Local Account variants.}
		\label{fig:time1}
	\end{subfigure}
	\hfill
	\begin{subfigure}[b]{\textwidth}
		\centering
		\includegraphics[width=\textwidth]{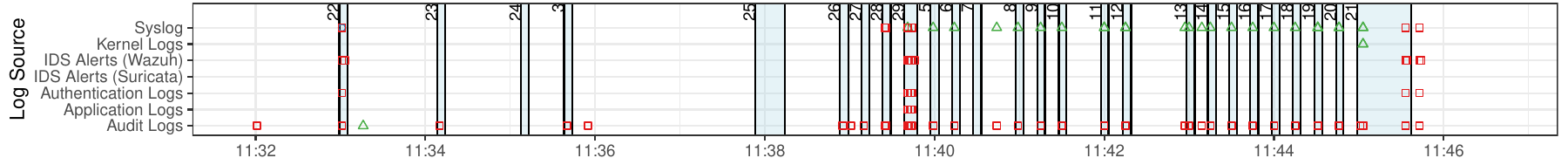}
		\caption{Scenario 2: Linux Malware with the Rootkit variant.}
		\label{fig:time2}
	\end{subfigure}
	\hfill
	\begin{subfigure}[b]{\textwidth}
		\centering
		\includegraphics[width=\textwidth]{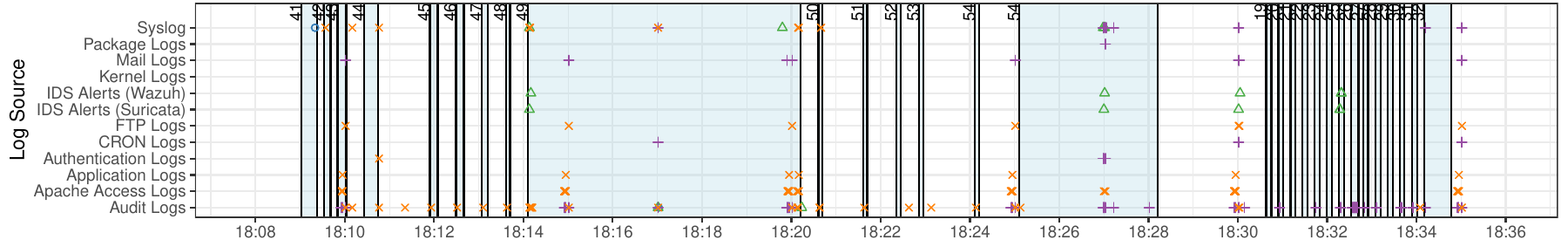}
		\caption{Scenario 3: Lateral Movement with the VNC and Apt variants.}
		\label{fig:time3}
	\end{subfigure}
	\hfill
	\begin{subfigure}[b]{\textwidth}
	\centering
	\includegraphics[width=\textwidth]{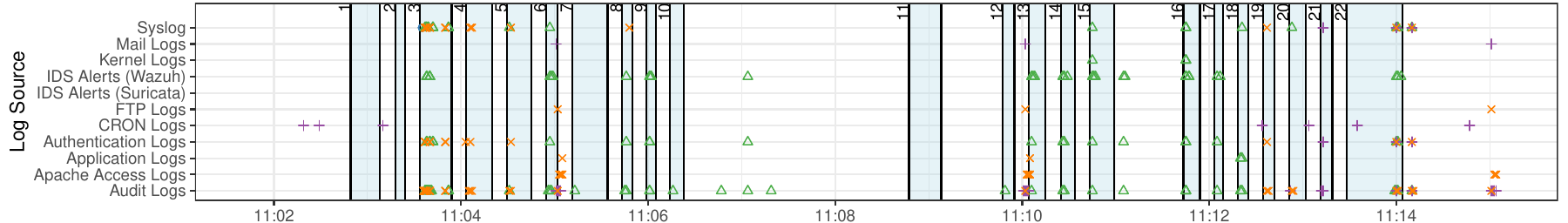}
	\caption{Scenario 4: Network Attack.}
	\label{fig:time4}
	\end{subfigure}
	\hfill
	\begin{subfigure}[b]{\textwidth}
	\centering
	\includegraphics[width=\textwidth]{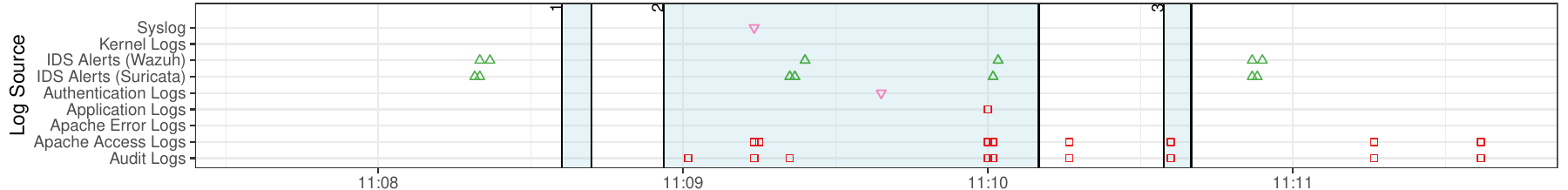}
	\caption{Scenario 5: Network Sniffing.}
	\label{fig:time5}
	\end{subfigure}
	\hfill
	\begin{subfigure}[b]{\textwidth}
	\centering
	\includegraphics[width=\textwidth]{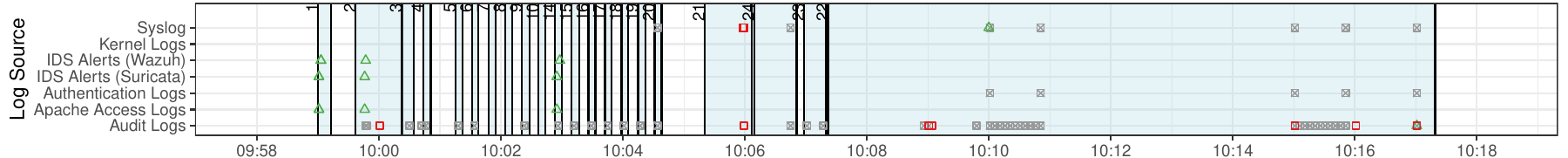}
	\caption{Scenario 6: Attack on Client with the Macro and CRON variants.}
	\label{fig:time6}
	\end{subfigure}
	\hfill
	\begin{subfigure}[b]{\textwidth}
	\centering
	\includegraphics[width=\textwidth]{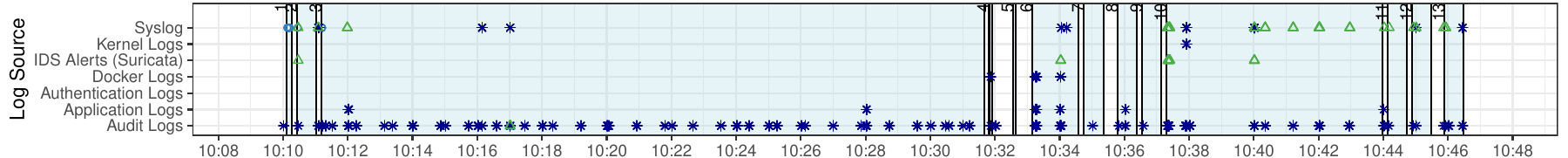}
	\caption{Scenario 7: Docker Attack.}
	\label{fig:time7}
	\end{subfigure}
	\caption{Timelines of log event and alert occurrences. Shaded intervals indicate active attack steps.}
	\label{fig:timelines}
\end{figure*}

Figure \ref{fig:timelines} visualizes the attack manifestations of Scenarios 1–7 based on the filtered data sets described in the previous section. For brevity, we display only a single attack path per scenario, using arbitrarily selected variants. Shaded intervals indicate the start and end times of individual attack steps; the numbers above these intervals correspond to the attack step identifiers used in Figs. \ref{fig:sc1}–\ref{fig:sc7}. Each point in the plots represents the occurrence of a log event or alert, where the vertical axis indicates the source of the event and the color denotes the host on which the event was generated.

The visualizations show that log events accumulate during specific attack steps, with some steps triggering events across multiple log sources or on several hosts. We consider this as an indicator that our filtering procedure successfully removes most events corresponding to normal background activity, while preserving events that are direct consequences of attacker behavior. Furthermore, the figures show that the majority of log events are generated on hosts that are the primary targets of the respective scenarios, such as the video server in Scenario 1 and the client machine in Scenario 6. A small number of events occur outside attack intervals; these correspond either to background activity that was not filtered out or to attack-related events associated with steps that are not explicitly labeled. The figure also illustrates that certain attack steps do not result in any observable log events at all. 

\subsection{Coverage of Attack Tactics and Techniques}

\begin{table}[t]
	\centering
	\scriptsize
	\caption{Number of attack techniques in scenarios}
	\label{tab:coverage}
\begin{tabular}{lccccccccc}
	Tactic & \rotatebox{90}{Scenario 1} & \rotatebox{90}{Scenario 2} & \rotatebox{90}{Scenario 3} & \rotatebox{90}{Scenario 4} & \rotatebox{90}{Scenario 5} & \rotatebox{90}{Scenario 6} & \rotatebox{90}{Scenario 7} & \rotatebox{90}{Total (Distinct)} & \rotatebox{90}{\makecell[l]{MITRE ATT\&CK \\ Coverage}} \\
	\toprule
	Reconnaissance & 4 & 0 & 0 & 0 & 0 & 0 & 4 & 5 & 45.5\% \\
	Resource Development & 0 & 0 & 0 & 0 & 0 & 0 & 0 & 0 & 0.0\% \\
	Initial Access & 2 & 0 & 2 & 1 & 0 & 1 & 3 & 4 & 36.4\% \\
	Execution & 2 & 1 & 3 & 0 & 0 & 3 & 3 & 6 & 35.3\% \\
	Persistence & 7 & 3 & 4 & 3 & 0 & 4 & 4 & 13 & 56.5\% \\
	Privilege Escalation & 7 & 3 & 3 & 3 & 0 & 3 & 2 & 9 & 64.3\% \\
	Defense Evasion & 6 & 4 & 4 & 6 & 1 & 4 & 2 & 15 & 31.9\% \\
	Credential Access & 2 & 1 & 3 & 0 & 3 & 1 & 2 & 7 & 41.2\% \\
	Discovery & 16 & 8 & 6 & 1 & 1 & 2 & 4 & 20 & 58.8\% \\
	Lateral Movement & 0 & 1 & 3 & 0 & 1 & 0 & 1 & 5 & 55.6\% \\
	Collection & 0 & 3 & 2 & 0 & 1 & 2 & 0 & 8 & 47.1\% \\
	Command and Control & 1 & 3 & 2 & 3 & 0 & 2 & 3 & 6 & 33.3\% \\
	Exfiltration & 0 & 1 & 0 & 0 & 0 & 0 & 1 & 1 & 11.1\% \\
	Impact & 0 & 0 & 6 & 0 & 0 & 0 & 0 & 6 & 40.0\% \\
	\midrule
	Total (Distinct) & 35 & 24 & 28 & 10 & 4 & 16 & 22 & 81 & 37.5\% \\
	\bottomrule
\end{tabular}
\end{table}

As outlined in Sect. \ref{scenarios}, the attack scenarios are designed to represent coherent kill chains that collectively cover a broad range of techniques from the MITRE ATT\&CK framework \cite{shen2024decoding}. Table \ref{tab:coverage} summarizes the number of distinct techniques present in each scenario, grouped by their corresponding tactics. Note that techniques associated with multiple tactics are counted separately for each of them. The table shows that all scenarios span multiple tactics, while also revealing that certain scenarios emphasize specific tactics, such as Discovery in Scenario 1 or Impact in Scenario 3.

In addition to the per-scenario breakdown, we also report the total number of distinct techniques covered across all scenarios for each tactic, excluding duplicate techniques that appear in multiple scenarios or are associated with multiple tactics. To contextualize these numbers, we report the corresponding coverage relative to the complete set of techniques defined in the MITRE ATT\&CK framework in the final column of the table.

Overall, the scenarios comprise a total of 81 distinct techniques, corresponding to 37.5\% of all techniques defined in the MITRE ATT\&CK framework and covering 16 out of the 19 most prevalent techniques mentioned across 667 cyber threat intelligence reports \cite{rahman2024attackers}. Several techniques are executed multiple times across different scenarios, using different methods and in varying contexts, which enables a broad analysis of their manifestations in log data. Specifically, 34 distinct techniques are part of at least two scenarios. Furthermore, some techniques are realized through multiple sub-techniques, which are labeled accordingly. When accounting these labels, the scenarios cover a total of 97 out of 475 distinct sub-techniques specified in MITRE ATT\&CK.

\subsection{Cyber Attack Manifestations} \label{manifestations}

This section analyzes how attack techniques manifest in system log data.

\subsubsection{Observability of Attack Commands} \label{commands}

\begin{figure*}
	\centering
	\includegraphics[width=\textwidth]{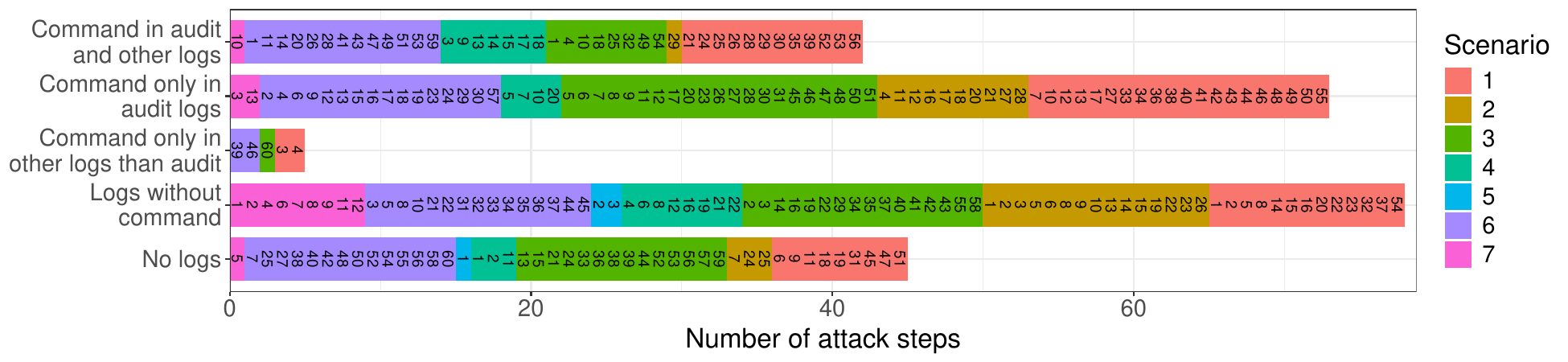}
	\caption{Observability of attacks and commands executed by the attacker in log data.}
	\label{fig:visibility}
\end{figure*}

Both manual log analysis by experts as well as LLM-based log interpretation rely heavily on the expressiveness of log event contents. Ideally, every action carried out by attackers generates log events that contain sufficiently granular information to make the underlying activity straightforward to understand. In practice, however, logs often capture only indirect consequences of actions rather than the actions themselves, or lack essential contextual information. To better understand how attacks manifest in log data, we investigate which attack techniques generate log events that contain explicit indicators of attacker activity.

To this end, we extract the commands executed by our attack emulation tool for each attack step and search the collected log data for occurrences of these commands. Specifically, we split each command into individual tokens on spaces and remove common stop-words (e.g., ``sudo'', ``id'') that yield too many matches in unrelated events. We consider an attack step to contain an explicit command indicator if at least one of the remaining tokens appears in any log events captured during the time interval where that attack step was captured.

Figure \ref{fig:visibility} summarizes the results of this experiment. Overall, a substantial fraction of attack steps generate log events that contain information about the executed command. Out of the 243 attack steps in our data set (cf. Sect. \ref{scenarios}), 115 (47.3\%) leave such indicators in audit logs. This result is largely explained by the fact that many attack chains involve direct command execution on compromised systems, for example as part of information gathering. Since audit logs record executed commands, they constitute a particularly rich source of information for log interpretation. Figure \ref{fig:cat1} shows sample audit log events capturing the command ``cat /etc/shadow'', which the attacker uses to dump credential hashes. Note that in some cases, audit logs encode command information in hexadecimal form; in the provided example, decoding the ``proctitle'' field also reveals the executed command.

Other log sources, including syslog, access logs, authentication logs, etc., also contribute relevant information. In addition to audit logs, 42 attack steps (17.3\%) leave explicit command indicators in other log sources, while 5 attack steps (2.1\%) generate such indicators exclusively outside of audit logs. The latter primarily involves remote tools that are not executed directly on the target system. For example, the access logs shown in Fig. \ref{fig:nikto} correspond to scanning activity and contain the keyword ``nikto'', identifying the scanner used by the attacker from outside the network. At the same time, our analysis shows that 78 attack steps (32.1\%) generate log events that do not directly reference the executed command, and 45 attack steps (18.5\%) that do not involve any log events at all after filtering.

\begin{figure*}[!tbhp]
\centering
\begin{subfigure}[b]{\textwidth}
\begin{lstlisting}[language=logs]
type=SYSCALL msg=audit(1765563116.496:15535): arch=c000003e syscall=59 success=yes exit=0 a0=5583f2d27de0 a1=5583f2e33fe0 a2=5583f2e2ff80 a3=8 items=2 ppid=4269 pid=4289 auid=4294967295 uid=0 gid=0 euid=0 suid=0 fsuid=0 egid=0 sgid=0 fsgid=0 tty=pts2 ses=4294967295 <!*comm="cat" exe="/usr/bin/cat"*!> subj=unconfined key="T1078_Valid_Accounts"
type=EXECVE msg=audit(1765563116.496:15535): argc=2 <!*a0="cat" a1="/etc/shadow"*!>
type=PATH msg=audit(1765563116.496:15535): item=0 <!*name="/usr/bin/cat"*!> inode=1568 dev=fc:01 mode=0100755 ouid=0 ogid=0 rdev=00:00 nametype=NORMAL cap_fp=0 cap_fi=0 cap_fe=0 cap_fver=0 cap_frootid=0
type=PATH msg=audit(1765563116.496:15535): item=1 name="/lib64/ld-linux-x86-64.so.2" inode=79850 dev=fc:01 mode=0100755 ouid=0 ogid=0 rdev=00:00 nametype=NORMAL cap_fp=0 cap_fi=0 cap_fe=0 cap_fver=0 cap_frootid=0
type=PROCTITLE msg=audit(1765563116.496:15535): <!*proctitle=636174002F6574632F736861646F77*!>
\end{lstlisting}
\caption{Command visible as clear text and hexadecimal-encoded text in audit logs of Scenario 3, Step 45.} 
\label{fig:cat1}
\end{subfigure}
\hfill
\begin{subfigure}[b]{\textwidth}
\begin{lstlisting}[language=logs]
type=SYSCALL msg=audit(1768992356.109:5653): arch=c000003e syscall=257 success=yes exit=7 a0=ffffffffffffff9c a1=c00046ec50 a2=80000 a3=0 items=1 ppid=3689 pid=3717 auid=0 uid=0 gid=0 euid=0 suid=0 fsuid=0 egid=0 sgid=0 fsgid=0 tty=(none) ses=31 comm="tool" exe="/opt/tool" subj=unconfined key="T1087_Account_Discovery"
type=PATH msg=audit(1768992356.109:5653): item=0 <!*name="/etc/shadow"*!> inode=5154 dev=fd:01 mode=0100640 ouid=0 ogid=42 rdev=00:00 nametype=NORMAL cap_fp=0 cap_fi=0 cap_fe=0 cap_fver=0 cap_frootid=0
type=PROCTITLE msg=audit(1768992356.109:5653): proctitle="/opt/tool"
\end{lstlisting}
\caption{Parts of command visible in audit logs of Scenario 7, Step 13.} 
\label{fig:cat2}
\end{subfigure}
\hfill
\begin{subfigure}[b]{\textwidth}
\begin{lstlisting}[language=logs]
Sep 22 18:58:45 videoserver sudo: webadmin : PWD=/home/webadmin ; USER=root ; <!*COMMAND=/usr/bin/cat /etc/shadow*!>
Sep 22 18:58:45 videoserver sudo: pam_unix(sudo:session): session opened for user root(uid=0) by (uid=1003)
Sep 22 18:58:45 videoserver sudo: pam_unix(sudo:session): session closed for user root
\end{lstlisting}
\caption{Command visible in authentication logs of Scenario 1, Step 24.} 
\label{fig:cat3}
\end{subfigure}
\caption{Observability of the attack command for credential dumping (T1003.008) across scenarios.}
\label{fig:cat}
\end{figure*}

\begin{figure*}[t]
\centering
\begin{lstlisting}[language=logs]
192.42.1.174 - - [22/Sep/2025:18:30:44 +0000] "<!*GET /old/*!> HTTP/1.1" 404 396 "-" "Mozilla/5.00 (<!*Nikto*!>/2.1.5)"
192.42.1.174 - - [22/Sep/2025:18:30:44 +0000] "<!*GET /oracle*!> HTTP/1.1" 404 396 "-" "Mozilla/5.00 (<!*Nikto*!>/2.1.5)"
192.42.1.174 - - [22/Sep/2025:18:30:44 +0000] "<!*GET /oradata/*!> HTTP/1.1" 404 396 "-" "Mozilla/5.00 (<!*Nikto*!>/2.1.5)"
192.42.1.174 - - [22/Sep/2025:18:30:44 +0000] "<!*GET /order/*!> HTTP/1.1" 404 396 "-" "Mozilla/5.00 (<!*Nikto*!>/2.1.5)"
192.42.1.174 - - [22/Sep/2025:18:30:44 +0000] "<!*GET /orders/*!> HTTP/1.1" 404 396 "-" "Mozilla/5.00 (<!*Nikto*!>/2.1.5)"
192.42.1.174 - - [22/Sep/2025:18:30:44 +0000] "<!*GET /orders/checks.txt*!> HTTP/1.1" 404 396 "-" "Mozilla/5.00 (<!*Nikto*!>/2.1.5)"
\end{lstlisting}
\caption{Excerpt of the log events generated as a consequence of scanning activities from Scenario 1, Step 3.} 
\label{fig:nikto}
\end{figure*}

Importantly, the presence of command strings in log data does not imply that an attack step is straightforward to interpret. First, identifying relevant commands becomes challenging when they are concealed by large volumes of log events with diverse parameters. Second, correct interpretation often depends on contextual information. For instance, audit logs generally record executions of commands such as ``chmod'', which are useful to make attack scripts executable (e.g., Scenario 6, Step 13); however, the same command may also be applied on benign files as part of routine administrative activity. Distinguishing benign from malicious use thus requires additional context, such as affected files and directories, the involved user, the temporal context of the action, etc.

A key finding of our analysis is that attack manifestations depend not only on which command is executed, but also on how it is executed. Figure \ref{fig:cat} illustrates this effect using three attack steps from different scenarios that all execute the same command: ``cat /etc/shadow''. In Scenario 3, the attacker executes the command via an interactive VNC session, resulting in audit logs that clearly expose both the command and its arguments (cf. Fig. \ref{fig:cat1}). In Scenario 7, the same command is executed through an implant, and the corresponding logs (cf. Fig. \ref{fig:cat2}) reveal only the accessed file path while obscuring the command itself. In Scenario 1, where the attacker interacts with the system through a reverse-shell, the command appears in authentication logs (cf. Fig. \ref{fig:cat3}). Since even identical commands may surface in different log sources and with varying levels of granularity depending on the execution method, we conclude that no consistent mapping between attack techniques and specific log manifestations can be established.

\subsubsection{Log Event Frequencies} \label{freq}

\begin{figure*}
	\centering
	\includegraphics[width=\textwidth]{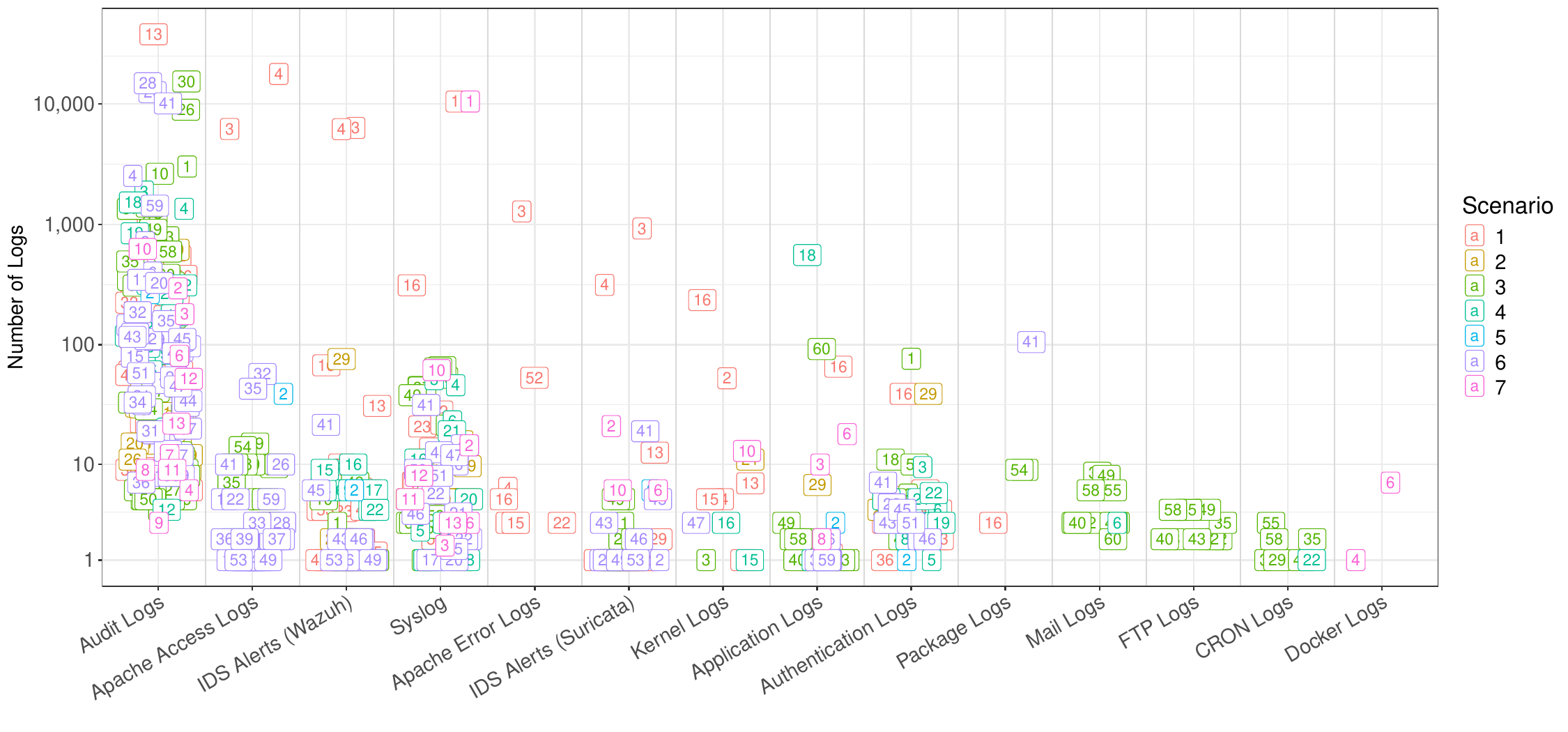}
	\caption{Log event frequencies observed in different log sources. The number in the tags reference the attack step of the respective scenario.}
	\label{fig:freq}
\end{figure*}

Aside from the contents of individual log messages, analysts frequently monitor the frequency of log events, for example, through dashboards that aggregate event counts over fixed time intervals. Changes in event frequencies are commonly used as basic indicators of anomalies in system behavior, because significant deviations such as spikes that stand out from long-term baselines may point to system issues or malicious activity \cite{he2016experience, landaueraminer}.

Several attack techniques included in our data set generate large volumes of log events. It is difficult to determine, however, which of these techniques would trigger suspicion in operational environments, because this kind of detection depends on the individual event count baselines. Since our data set does not include simulations of benign workload behavior, we only compare the frequencies of log events generated by different attack steps with each other. Figure \ref{fig:freq} visualizes the number of log events produced per attack step across log sources. As shown in the figure, audit logs account for the largest share of events. The main reason for this is that a large number of attack steps directly interact with systems and thus trigger audit logging, causing anything from a few to several thousands of events. Thereby, the system vulnerability scan performed in Step 13 of Scenario 1 produces the highest volume of audit logs, generating more than 40,000 events.

In contrast, most other log sources typically do not generate more than 100 events per attack step, often substantially fewer. Notable exceptions to this tendency are attack steps involving network-based scanning activities. For example, the DNS enumeration in Step 1 of Scenario 1 results in a high number of syslog entries on the DNS server, the network scan in Step 2 of Scenario 1 generates numerous system logs in the firewall, and the network vulnerability scan in Step 3 of Scenario 1 produces a large volume of access logs. An excerpt of the latter is shown in Fig. \ref{fig:nikto}, which depicts a rapid succession of requests corresponding to entries from a wordlist. Similar log characteristics can be observed for the brute-force login executed in Step 1 of Scenario 3. Figure \ref{lst:auth} shows an excerpt of the corresponding logs that document a series of failed authentication events taking place within a few seconds.

\begin{figure*}[t]
\centering
\begin{lstlisting}[language=logs]
Dec 11 14:28:03 puppet sshd[3089]: <!*Invalid user admin*!> from 192.42.1.174 port 33006
Dec 11 14:28:03 puppet sshd[3089]: Received disconnect from 192.42.1.174 port 33006:11: Bye Bye [preauth]
Dec 11 14:28:03 puppet sshd[3095]: pam_unix(sshd:auth): check pass; <!*user unknown*!>
Dec 11 14:28:03 puppet sshd[3095]: pam_unix(sshd:auth): <!*authentication failure*!>; logname= uid=0 euid=0 tty=ssh ruser= rhost=192.42.1.174
Dec 11 14:28:05 puppet sshd[3095]: <!*Failed password for invalid user admin*!> from 192.42.1.174 port 56768 ssh2
\end{lstlisting}
\caption{Failed authentication events generated as a consequence of a brute-force login from Scenario 3, Step 1.} 
\label{lst:auth}
\end{figure*}

\subsubsection{System Performance Metrics}

\begin{figure}[t]
\centering
\begin{lstlisting}[language=logs]
[2025-09-22 18:37:00] [info] write_log values:
videoserver_novalocal.load.<!*load.shortterm 0.1*!>
[2025-09-22 18:37:10] [info] write_log values:
videoserver_novalocal.load.<!*load.shortterm 0.24*!>
[2025-09-22 18:37:20] [info] write_log values:
videoserver_novalocal.load.<!*load.shortterm 1.12*!>
[2025-09-22 18:37:30] [info] write_log values:
videoserver_novalocal.load.<!*load.shortterm 1.71*!>
[2025-09-22 18:37:40] [info] write_log values:
videoserver_novalocal.load.<!*load.shortterm 1.44*!>
\end{lstlisting}
\caption{System metric captured with collectd.} 
\label{lst:metrics}
\end{figure}

Log collection frameworks such as collectd\footnote{\url{https://collectd.org/}} capture system performance metrics over time. The collected measurements provide insights into system utilization at given points in time and are particularly useful for correlating system activities with changes in system performance, for example, when programs with high computational workload are executed. Common categories of metrics include CPU, file system, disk, entropy, network interfaces, interrupt counters, load, memory, processes, and swap usage; each category may comprise multiple individual measurements. Figure \ref{lst:metrics} shows an excerpt of collectd logs capturing short-term system load in intervals of ten seconds. The displayed measurements indicate a noticeable increase in system load, which is caused by the system vulnerability scan (Scenario 1, Step 13).

\begin{figure*}
	\centering
	\includegraphics[width=\textwidth]{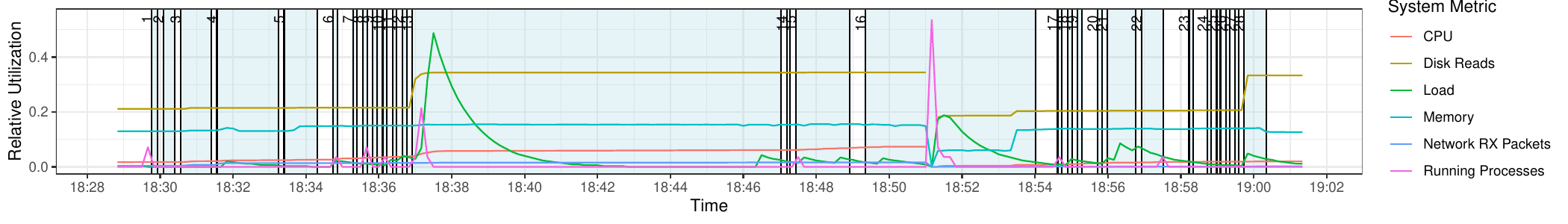}
	\caption{Timelines of system performance metrics collected from the video server in Scenario 1 (Autostart, Local Account).} 
	\label{fig:metrics}
\end{figure*}

Figure \ref{fig:metrics} visualizes the evolution of system performance metrics captured during the entire simulation period of Scenario 1, including system load, CPU utilization of user-space processes, the number of disk read operations, total memory usage, the number of received network packets, and the number of running processes. Note that for visualization purposes, all metrics are scaled to a common range. The figure reveals that certain attack steps have a measurable impact on system performance. Corresponding to the logs displayed in Fig. \ref{lst:metrics}, the system vulnerability scan executed in Step 13 leads to a sharp increase in CPU utilization, disk read activity, and system load. These results demonstrate that attack steps do not only manifest through generation of log events but also impact measurements in continuously generated system performance metrics.

\subsubsection{Intrusion Detection Alerts} \label{alerts}

Certain types of log events are considered suspicious because they are commonly produced as consequences of malicious or otherwise undesired system activities. Similarly, specific behavioral patterns observable in network traffic can indicate potential attacks. Intrusion Detection Systems (IDS) analyze system logs and network traffic for matches against predefined detection rules and signatures, triggering alerts when such patterns are observed. As described in Sect. \ref{collection}, our experimental setup includes the host-based IDS Wazuh and the network-based IDS Suricata, both configured to use their respective publicly available default rule sets.

\begin{figure}[t]
\centering
\begin{lstlisting}[language=logs]
{
  "rule": {
    "level": 10,
    "description": "<!*Auditd: Device enables promiscuous mode.*!>",
    "id": "80710"
  },
  "full_log": "<!*type=ANOM_PROMISCUOUS*!> msg=audit(1765797045.372:6848): dev=ens3 prom=256 old_prom=0 auid=4294967295 uid=0 gid=0 ses=4294967295",
  "location": "/var/log/audit/audit.log"
}
{
  "rule": {
    "level": 3,
    "description": "<!*Successful sudo to ROOT executed.*!>",
    "id": "5402"
  },
  "full_log": "Dec 15 11:10:45 inetfw sudo: root : TTY=pts/1 ; PWD=/usr/bin ; USER=root ; <!*COMMAND=/usr/bin/systemctl start auditf.service*!>",
  "location": "/var/log/auth.log"
}
\end{lstlisting}
\caption{Alerts triggered by Wazuh IDS when installing the port knocking tool in Scenario 4, Step 15.} 
\label{lst:wazuh}
\end{figure}

\begin{figure*}[t]
\centering
\begin{lstlisting}[language=logs]
12/11/2025-14:42:48.133080  [**] [1:2034567:1] ET HUNTING <!*curl User-Agent to Dotted Quad*!> [**] [Classification: Potentially Bad Traffic] [Priority: 2] {TCP} 192.168.100.23:49346 -> 192.42.1.174:80
12/11/2025-14:42:48.133803  [**] [1:2019240:14] ET POLICY <!*Executable and linking format (ELF) file download Over HTTP*!> [**] [Classification: Potential Corporate Privacy Violation] [Priority: 1] {TCP} 192.42.1.174:80 -> 192.168.100.23:49346
\end{lstlisting}
\caption{Alerts triggered by Suricata IDS when downloading ransomware in course of Scenario 3, Step 25.} 
\label{lst:suricata}
\end{figure*}

Figures \ref{lst:wazuh} and \ref{lst:suricata} illustrate sample alerts generated by these systems. Figure \ref{lst:wazuh} shows alerts triggered by Wazuh during Step 15 of Scenario 4, in which the attacker installs a port-knocking tool. The first alert is a high-severity alert (level 10) indicating that a network interface has been switched into promiscuous mode, which is typically associated with network monitoring and therefore considered security-critical. The second alert has a substantially lower severity (level 3) and records the execution of a privileged command. The main reason for the low severity is that these commands are common in benign administrative workflows. Figure \ref{lst:suricata} presents two alerts generated by Suricata during Step 25 of Scenario 3, in which the attacker downloads ransomware. The first alert has medium severity (priority 2) and flags an HTTP request to an IP address, which is in our case controlled by the attacker. The second alert has the highest severity level (priority 1) and is triggered because an executable file is downloaded over unencrypted HTTP. Note that Suricata and Wazuh use different severity scales, which necessitates normalization for comparative analysis.

\begin{figure}
	\centering
	\includegraphics[width=.49\textwidth]{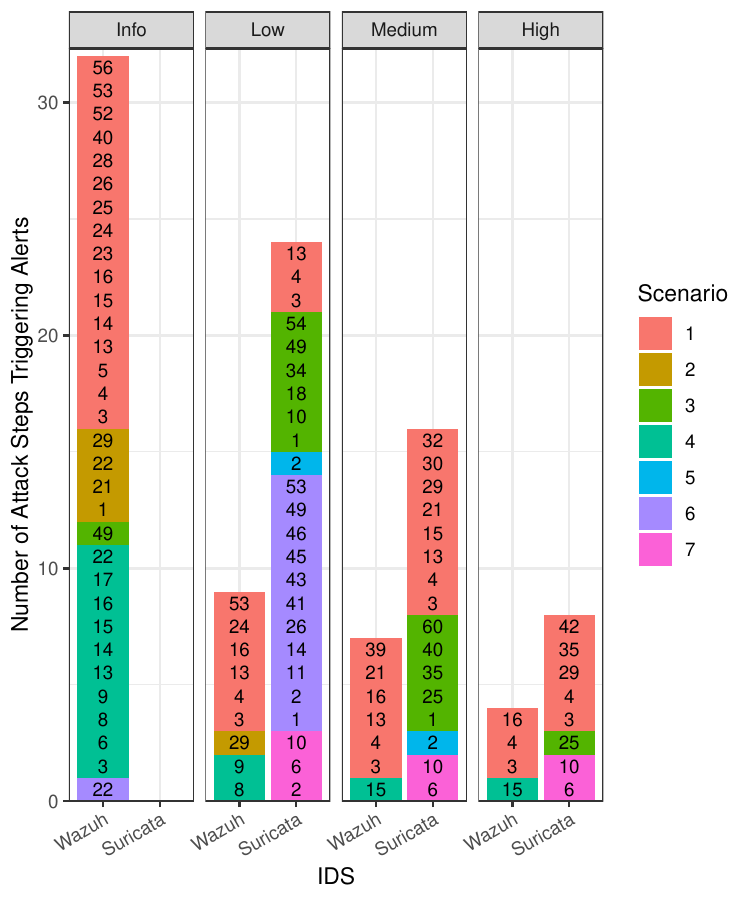}
	\caption{Severities of intrusion detection alerts across scenarios and attack steps.}
	\label{fig:severity}
\end{figure}

We analyze all IDS alerts triggered during the execution of our attack scenarios. Figure \ref{fig:severity} summarizes the attack steps that result in alerts and groups them by severity. To enable comparability across IDS, we map Wazuh’s severity levels onto the three-point scale used by Suricata as follows: levels 4–6 correspond to Low, levels 7–9 to Medium, and level 10 and above to High. Wazuh alerts with severity of level 3 or lower are treated as Informational, as they do not necessarily indicate malicious activity. Note that single attack steps may trigger multiple alerts across different severity levels (cf. Figs. \ref{lst:wazuh} and \ref{lst:suricata}) and from both IDSs at the same time. For example, the scanning activities in Steps 3 and 4 of Scenario 1 generate alerts from both Wazuh and Suricata, spanning all severity categories. Overall, the results show that Suricata triggers alerts for a larger number of attack steps in the Low-High severity ranges, whereas Wazuh predominantly generates informational alerts across many steps.

Our analysis shows that out of the 198 attack steps that produce log data, only 43 trigger alerts with a severity level of Low-High in at least one of the two IDS. An additional 20 attack steps trigger informational alerts in Wazuh. Conversely, 155 attack steps remain undetected by either IDS when relying on default detection signatures. These results highlight the limitations of signature-based intrusion detection in capturing a wide range of attacker activities and underscore the importance of complementary analysis approaches, such as methods for anomaly-based detection or LLM-based log interpretation.

\section{LLM-based Interpretation} \label{interpretation}

This section presents an illustrative evaluation of our data set in the context of log interpretation using LLMs.

\subsection{Experimental Setup}

In this section, we summarize the setup of our experiment.

\subsubsection{Task \& Model}

We task an LLM with interpreting log data in a zero-shot setting, i.e., without providing any training samples. The goal of this experiment is to approximate a realistic usage scenario in which security analysts copy and paste log excerpts into an LLM to obtain a fast and automated first assessment that complements manual analysis \cite{singh2025llms}. For this experiment, we use ChatGPT\footnote{We use ChatGPT version 5.2, which is the most recent version available at the time of conducting this experiment. }, a model that has also been employed for log interpretation in prior work \cite{qi2023loggpt, liu2024logprompt}. Moreover, we set the model's temperature to zero to minimize hallucinations.

\subsubsection{Prompting}

For each of the 198 attack steps that generate log data, we submit one prompt to ChatGPT. Thereby, each prompt starts with a task description instructing the model to (i) classify the provided log data according to the MITRE ATT\&CK framework and return the top 10 most relevant techniques in descending order, (ii) estimate whether the logs correspond to malicious or benign behavior on a 7-point likert scale, and (iii) provide a short explanation on the reasoning behind the assessment.

\begin{figure*}[!tbhp]
	\centering
	\begin{lstlisting}[language=logs]
You are a MITRE ATT&CK TTP classification expert. Your task is to classify the following system log data. You are provided with samples from one or more hosts and one or more log sources that are captured during execution of one specific MITRE ATT&CK technique. Always output a valid JSON object with the following fields:
 - "techniques": A list of top 10 ATT&CK techniques that are most likely related to the sample logs, sorted in descending order. Only print the ID of the techniques without any other descriptions.
 - "confidence": An estimate for the certainty that the logs indicate an actual attack rather than normal system or user activity. Provide one of the following estimates: "Certain: Attack", "Almost Certain: Attack", "Somewhat Certain: Attack", "Neutral", "Somewhat Certain: Normal", "Almost Certain: Normal", "Certain: Normal"
 - "explanation": A brief explanation (1-2 sentences) why you think that the samples correspond to attacks or normal behavior, e.g., by pointing to specific artifacts or properties of the logs.

videoserver:
  logs/access.log.1:
    <!*192.42.1.174*!> - - [24/Sep/2025:11:31:05 +0000] "<!*GET /zm/index.php*!> HTTP/1.1" 200 8978 "-" "Mozilla/5.0 (Windows NT 10.0; Win64; x64; rv:133.0) Gecko/20100101 Firefox/133.0"
    <!*192.42.1.174*!> - - [24/Sep/2025:11:31:05 +0000] "<!*POST /zm/index.php*!> HTTP/1.1" 302 756 "-" "Mozilla/5.0 (Windows NT 10.0; Win64; x64; rv:133.0) Gecko/20100101 Firefox/133.0"
  logs/log/zm/web_php.log:
    9/24/25, 11:31:12 AM UTC.759462 web_php[2639].WAR [<!*192.42.1.174*!>] [<!*No event returned from TriggerOn*!>] at /usr/share/zoneminder/www/includes/Monitor.php line 678
  logs/log/audit/audit.log:
    type=SYSCALL msg=audit(1758713480.172:4453): arch=c000003e syscall=59 success=yes exit=0 a0=55df1de3c000 a1=55df1de3bf88 a2=55df1de3bfa0 a3=8 items=4 ppid=2688 pid=2689 auid=4294967295 <!*uid=33 gid=33*!> euid=33 suid=33 fsuid=33 egid=33 sgid=33 fsgid=33 tty=(none) ses=4294967295 comm="which" exe="/usr/bin/dash" subj==unconfined key="T1166_Seuid_and_Setgid" ARCH=x86_64 SYSCALL=execve AUID="unset" <!*UID="www-data" GID="www-data"*!> EUID="www-data" SUID="www-data" FSUID="www-data" EGID="www-data" SGID="www-data" FSGID="www-data"
    type=EXECVE msg=audit(1758713480.172:4453): argc=3 <!*a0="/bin/sh" a1="/usr/bin/which" a2="python"*!>
	\end{lstlisting}
	\caption{Exemplary LLM prompt showing a task description and logs from Scenario 1, Step 5.}
\label{lst:prompt}
\end{figure*}

Following this task description, we include log data collected from all hosts and log sources involved in the respective attack step. To avoid excessive prompt lengths, we limit the number of logs per host–source pair at ten lines. These lines are sampled at random positions within the corresponding log file. We repeat this process five times per attack step with new random samples of log lines in order to mitigate the effects of sampling randomness and to assess the consistency of resulting classifications. Figure \ref{lst:prompt} shows an exemplary prompt, including both the instruction preamble and selected access and audit logs from the video server during the ZoneMinder exploit (Scenario 1, Step 5). The logs depict the request to the web server that transmits the malicious payload, a part of the subsequent execution of the payload on the system itself, as well as a warning message about a missing event from the ZoneMinder application. Note that we simplify some logs and remove less relevant log sources for brevity.

\begin{figure*}[t]
	\centering
	\begin{subfigure}[b]{\textwidth}
		\begin{lstlisting}[language=logs]
"techniques": ["<!*T1059*!>.004", <!*"T1059.006"*!>, "T1082", "T1518.001", "T1046", <!*"T1190"*!>, "T1105", "T1505.003", "T1078", "T1036"],
"confidence": "Somewhat Certain: Attack",
"explanation": "A web request to ZoneMinder (/zm/index.php) is followed by audit logs showing the web server account (uid/gid 33, www-data) spawning a shell to run command discovery (\"/bin/sh /usr/bin/which python\"), which is atypical for normal web browsing and suggests command execution via a web application. The ZoneMinder warning tied to the same source IP and the process execution from a web context increases likelihood of exploitation of a public-facing app."
		\end{lstlisting}
		\caption{ZoneMinder exploit in Scenario 1, Step 5.} 
		\label{fig:llm1}
	\end{subfigure}
	\hfill
	\begin{subfigure}[b]{\textwidth}
		\begin{lstlisting}[language=logs]
"techniques": ["T1548.003", <!*"T1003.008"*!>, "T1078", "T1059.004", "T1087.001", "T1106", "T1068", "T1555", "T1552.001", "T1036"],
"confidence": "Almost Certain: Attack",
"explanation": "Logs show user \"debian\" invoking sudo to run \"cat /etc/shadow\" as root, which is a common precursor to credential dumping (reading local password hashes) and matches Wazuh's mapping to sudo-based privilege escalation (T1548.003). Access to /etc/shadow via elevated privileges is rarely needed in routine operations and strongly indicates malicious credential access activity."
		\end{lstlisting}
		\caption{Credential dumping in Scenario 1, Step 24.} 
		\label{fig:llm2}
	\end{subfigure}
		\hfill
	\begin{subfigure}[b]{\textwidth}
		\begin{lstlisting}[language=logs]
"techniques": [<!*"T1595.002"*!>, "T1190", "T1083", "T1590.001", "T1590.004", "T1592.002", "T1213", "T1110", "T1505.003", "T1070.004"],
"confidence": "Almost Certain: Attack",
"explanation": "The logs show automated web probing with Nikto user-agent and many 404/invalid-URI requests for common/admin paths and CGI scripts, plus explicit traversal/LFI-style attempts to access sensitive files like /etc/passwd, boot.ini, and /system32/, consistent with vulnerability scanning and exploitation attempts against a public-facing web server."
		\end{lstlisting}
		\caption{Wordlist scanning in Scenario 1, Step 3.} 
		\label{fig:llm3}
	\end{subfigure}
	\caption{Samples of correct LLM-based classification and interpretation of logs from attack steps.}
	\label{fig:llm}
\end{figure*}

\subsection{Results}

This setup presents the results obtained from our experiment.

\subsubsection{Qualitative Examples}

Figure \ref{fig:llm} showcases responses for attack steps that are correctly interpreted. Figure \ref{fig:llm1} shows the response to the prompt in Fig. \ref{lst:prompt}. Based on the explanation generated by the LLM, it becomes evident that the model correctly connects the web request observed in the access logs with the command executed by the web application process recorded in the audit logs through temporal correlation, and identifies this behavior as unusual. Furthermore, the model links the ZoneMinder application warning message via IP-based correlation and interprets the event as corroborating evidence of potentially malicious activity. This form of multi-source reasoning resembles the analytical process expected from a human security analyst. The ground truth labels of the corresponding attack step, namely \textit{Exploitation of Public-Facing Application (T1190)} as well as \textit{Command and Scripting Interpreter (T1059)}, appear in the first and sixth positions of the top predictions, indicating a strong semantic understanding of both the meaning of the log content and the associated attack techniques. Figure \ref{fig:llm2} demonstrates the interpretation of credential dumping; relevant logs are depicted in Fig. \ref{fig:cat3}. The response shows that the correct technique and sub-technique (\textit{T1003.008}) appear at position two in the ranked list of predictions. The prediction for the top-ranked technique is likely influenced by the presence of a Wazuh alert referencing \textit{Sudo Caching for Privilege Escalation (T1548.003)}. Again, the explanation of the underlying cause of the logs and their relevance to specific malicious activities are on point. Figure \ref{fig:llm3} shows the interpretation of vulnerability scanning, with log excerpts shown in Fig. \ref{fig:nikto}. In this case, the LLM correctly identifies the scanning tool via the user-agent string as well as the high number of requests, thus leveraging both event content (cf. Sect. \ref{commands}) and frequency (cf. Sect. \ref{freq}). Once more, the top prediction matches the ground truth sub-technique \textit{Active Scanning: Vulnerability Scanning} (T1595.002).

\subsubsection{Quantitative Evaluation Results}

We evaluate classification performance by comparing the predicted techniques with the ground truth labels across all five runs of the 198 attack steps. Thereby, we only consider matches at the technique level since the model was not instructed to return sub-technique identifiers. We record the rank of the highest correct prediction, where any of the ground-truth techniques appears in the top 10 list.

\begin{figure*}
	\centering
	\includegraphics[width=\textwidth]{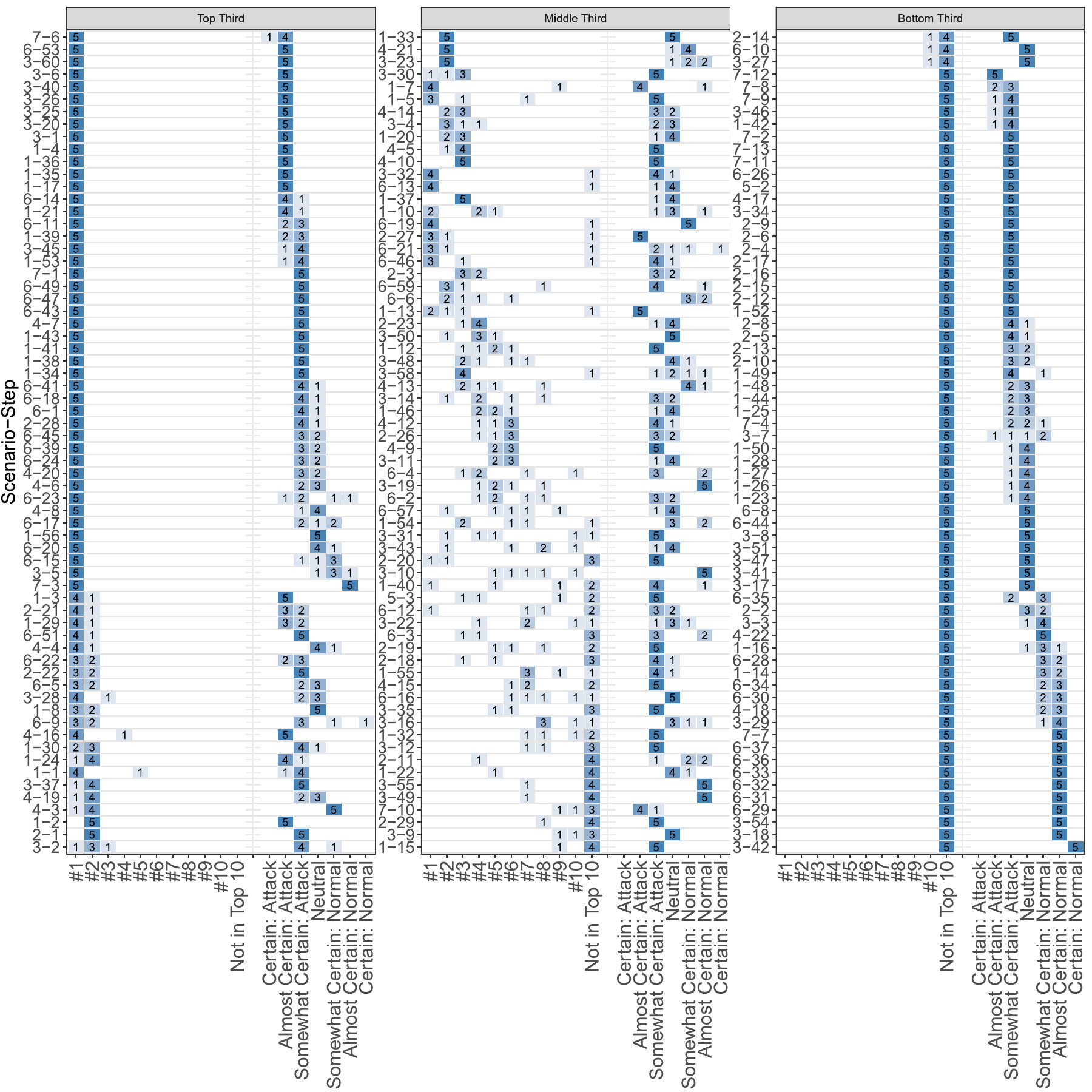}
	\caption{Presence of ground-truth labels among the top-10 predicted attack techniques and estimation of benign versus malicious origin.}
	\label{fig:llmclass}
\end{figure*}

Figure \ref{fig:llmclass} visualizes the results, showing both the rank of the highest matching technique (positions \#1–\#10 or not in top 10) and the corresponding likert-scale confidence estimates. For visualization purposes, we split the results into the top, middle, and bottom thirds with respect to attack step classification accuracy. The figure shows that the top third of the attack steps are classified with high accuracy, yielding correct technique predictions almost always at position \#1 or \#2. The middle third exhibits adequate results where a matching technique is found within the top 10 predictions. The bottom third contains hardly any correct predictions within the top 10. 

Confidence estimates regarding maliciousness vary substantially across attack steps and do not always correlate with classification accuracy. We observe no tendency for attack steps from particular scenarios to be easier or harder to classify than others, indicating that all scenarios are equally suitable for this type of evaluation.

\subsubsection{Impact of Manifestation Characteristics}

Finally, we investigate the relations between different types of attack manifestations and the LLM's classification performance. To this end, we correlate these top, middle, and bottom third of classification results (cf. Fig. \ref{fig:llmclass}) with previously analyzed manifestation characteristics, namely command observability (cf. Sect. \ref{commands}), event frequency (cf. Sect. \ref{freq}), and presence of IDS alerts (cf. Sect. \ref{alerts}).

\begin{figure*}[t]
	\centering
	\begin{subfigure}[t]{.27\textwidth}
		\centering
		\includegraphics[width=\textwidth]{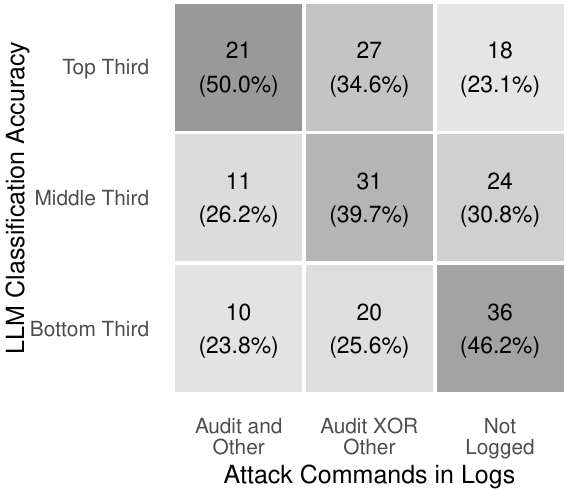}
		\caption{Observability of commands.}
		\label{fig:conf1}
	\end{subfigure}
	\quad
	\begin{subfigure}[t]{.27\textwidth}
		\centering
		\includegraphics[width=\textwidth]{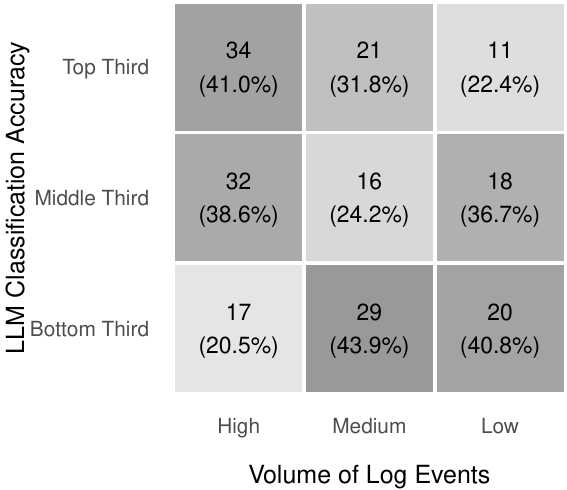}
		\caption{Log event frequencies.}
		\label{fig:conf2}
	\end{subfigure}
	\quad
	\begin{subfigure}[t]{.338\textwidth}
		\centering
		\includegraphics[width=\textwidth]{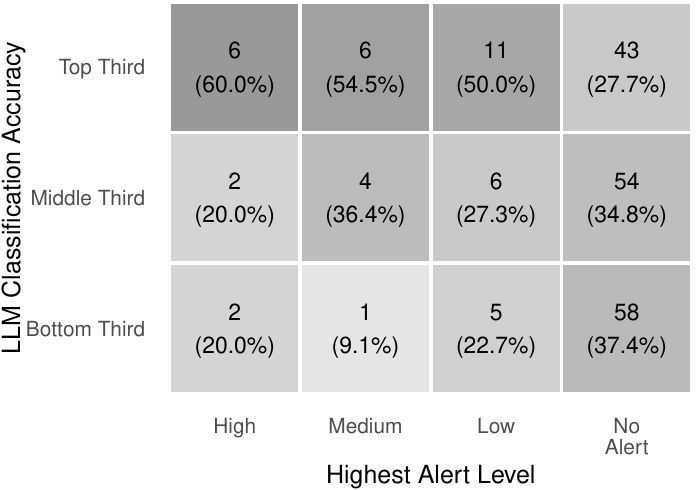}
		\caption{Highest alert severities.}
		\label{fig:conf3}
	\end{subfigure}
	\caption{Contingency tables of LLM classification accuracies and attack manifestations.}
	\label{fig:conf}
\end{figure*}

Figure \ref{fig:conf} presents the results in the form of contingency tables. Figure \ref{fig:conf1} shows that classification accuracy is highest when attack commands are visible in both audit logs and other log sources, lower when commands appear in only one of these sources, and lowest when no command information is present in the logs analyzed by the LLM. Figure \ref{fig:conf2} indicates that attack steps generating large numbers of log events ($>1000$ in audit or $>100$ in other log sources) are more accurately classified than those with moderate ($>100$ in audit or $>10$ in other log sources) or lower event volumes. This may be attributable to the presence of stronger indicators in high-frequency attacks (cf. Fig. \ref{fig:nikto}), a higher likelihood of relevant logs being sampled into the prompt, or the informative nature of frequency patterns themselves. Finally, Fig. \ref{fig:conf3} shows that most attack steps triggering IDS alerts fall into the top third of classification accuracy. This can be attributed either to the additional semantic information contained in the alerts themselves or to the fact that these attack steps are inherently more straightforward to identify as such and thus easier to interpret.

\section{Discussion} \label{discussion}

In this section, we discuss the implications of our findings. We subsequently outline limitations of our work and derive recommendations for future research.

\subsection{Implications}

This paper introduces a new data set of cyber attack manifestations in system log data and provides an empirical analysis of their characteristics. Our results indicate that the majority of attack steps leave distinct traces in system logs that allow to reconstruct individual actions. Our analysis suggests that attack executions can be broadly grouped into at least four types of manifestations. First, attacks that leave no observable manifestations in common sources for system log data, which may occur for certain types of interactions as well as for preparatory steps that take place outside the monitored systems or network. Second, direct manifestations that expose executed commands. Across most scenarios, information about executed commands is present in the logs, despite the fact that attackers employ different execution mechanisms, including SSH (Scenarios 1, 3, and 4), implants (Scenarios 2 and 4), reverse-shells (Scenarios 1 and 3), VNC (Scenario 3), and screen sharing software (Scenario 6). While these mechanisms differ substantially in how attackers interact with compromised systems, our examples show that command information is often still observable. At the same time, the execution method strongly influences the granularity and visibility of these traces, with certain approaches effectively obscuring command-level details. Third, manifestations that do not reveal executed commands but are generated as direct consequences of malicious actions. Although less explicit, such events can still serve as strong indicators of adversary behavior, particularly when they stand out due to unusual parameters (e.g., events indicating system configuration changes) or high event frequencies (e.g., scanning activities). Fourth, indirect manifestations, such as changes in system performance metrics. These manifestations are generally more difficult to relate to specific attack techniques; nonetheless, unexpected changes of these metrics may be useful to detect undesired activities that are not captured or detectable in other log data. 

Our results further show that while intrusion detection systems successfully trigger alerts for some attack steps, they fail to detect a large fraction of them. We acknowledge that many attack commands closely resemble legitimate administrative actions, and that IDS rule sets intentionally avoid flagging such activities to limit false positives. Nonetheless, this observation highlights the limitations of purely signature-based detection and suggests the need for complementary analysis approaches.

One such complementary approach is LLM-based log interpretation. In our illustrative evaluation, the LLM classified approximately one third of attack steps with near-perfect accuracy and another third within the top-10 technique predictions. This result is particularly notable given the large search space of 216 MITRE ATT\&CK techniques and the zero-shot setting, in which the model is provided with no contextual information beyond the log data generated during each attack step. These findings suggest that LLMs are in fact capable of extracting meaningful semantic patterns from system logs and have the potential to support automated log interpretation, thereby assisting human analysts in understanding attacker behavior.

\subsection{Limitations and Threats to Validity}

We acknowledge several limitations and threats to validity of our work. Our infrastructure for data collection is limited to Linux-based systems and open-source software, which we deliberately chose to ensure full reproducibility. However, attack manifestations may differ substantially on other operating systems such as Windows, where logging mechanisms, formats, and granularity, vary considerably. As a result, the performance of LLM-based log interpretation may differ significantly. Researchers should therefore complement our data set with Windows-based data sets such as the ones reviewed in Sect. \ref{datasets} for evaluations.

Even though our methodology relies on idle systems and log filtering (cf. Sects. \ref{procedure} and \ref{filtering}), we cannot guarantee that the collected data sets are entirely free of background noise. At the same time, our filtering procedure may remove events that are influenced by attacks. Some label inaccuracies may be introduced by delayed events that fall outside of their corresponding attack intervals or coincide with other attack steps. To mitigate these issues, we provide both filtered and unfiltered versions of the extracted attack manifestations (cf. Sect. \ref{intro}). Moreover, despite independent manual validation of attack playbooks and the resulting attack manifestations, we cannot guarantee that every attack step completed as expected since not all steps produce explicit output.

We also emphasize that our evaluation of LLM-based log interpretation is intended as an illustrative case study rather than a comprehensive benchmark. We consider only a single LLM and acknowledge that the observed results may not generalize to other models. Furthermore, limiting the prompt to ten randomly sampled events per log file means that the most informative events may not always be included, thereby affecting classification accuracy. Moreover, our evaluation strategy that counts correct classifications if any ground truth technique appears in the predicted list may bias results in favor of attack steps with many assigned techniques. For instance, system vulnerability scanning (Scenario 1, Step 13) is associated with 13 techniques, which increases the likelihood of obtaining at least one match in comparison to most attack steps which only involve one or two associated techniques.

Finally, our attack scenarios are not derived from real-world incidents but are intentionally constructed to combine as many attack techniques as practically feasible into coherent kill chains. While scenario design was carried out by domain experts with experience in offensive technologies and attack modeling, these sequences may not fully reflect real attacker behavior. In addition, despite independent cross-validation, the fact that the same team designed and labeled the scenarios may have introduced unintentional bias.

\subsection{Future Work}

Since the modeled attack cases are labeled at the technique level, the presented data set is directly applicable for evaluation of log interpretation approaches as demonstrated in our case study. Extending these labels could prove useful for evaluation of additional tasks for LLMs in the security domain, including log summarization \cite{mudgal2023assessment, liu2025loglm}. In particular, we assign natural-language descriptions to selected attack steps (cf. Fig. \ref{fig:playbook}), which could serve as alternative or complementary ground truth labels. Future extensions could also incorporate labels related to mitigation strategies and recommended countermeasures, e.g., commands to remove backdoors and return the system into a secure state. We plan to release enriched versions of the data set with extended annotations as part of future publications.

In our illustrative evaluation, we only consider raw log data and IDS alerts as input to the LLM. However, additional contextual information could substantially improve attack interpretation and classification. Specifically, our data set includes system configuration files collected from all hosts that contain details about installed software versions and system settings. Such asset information can provide valuable context that verifies or augments attack interpretations. For example, in the ZoneMinder exploit (Scenario 1, Step 5), the configuration files specify the exact version of the ZoneMinder application\footnote{We deploy ZoneMinder version 1.36.31 in our test environment.}, which is vulnerable to unauthenticated remote code execution according to public threat intelligence\footnote{All ZoneMinder versions prior to 1.36.33 and 1.37.33 are vulnerable to the exploit considered in Scenario 1: \url{https://nvd.nist.gov/vuln/detail/CVE-2023-26035}.}. While the LLM is already capable of correctly interpreting the logs generated as consequences of this exploit (cf. Fig. \ref{lst:prompt}) as command execution via the ZoneMinder web application (cf. Fig. \ref{fig:llm1}), the combination with asset data and external threat intelligence could further corroborate LLM interpretations, increase prediction confidence, enrich explanations with relevant vulnerability references, and potentially suggest actionable mitigation steps.

Another simplification of our evaluation is that we consider attack steps in isolation. In practice, correlating information across multiple steps would enable reasoning about the attacker’s objectives and the overall progression of the kill chain. Instead of submitting independent prompts for each step, future work could incorporate log data or classifications from preceding attack steps as contextual information. This could resolve incorrect interpretations regarding benign or malicious intentions of underlying actions (cf. Fig. \ref{fig:llmclass}). For example, commands executed by the attacker that are also commonly carried out by system administrators may appear as benign when considered in isolation, but are more likely to be interpreted correctly when immediately following other clearly malicious activities carried out by the same user or within the same context.

Furthermore, we use only a single general-purpose LLM in this paper but recommend to compare performance across different models and architectures in future evaluations. In particular, prior work has shown that LLM fine-tuning can improve log interpretation capabilities \cite{ji2025adapting}. Aside from that, given the already promising results obtained with a general-purpose model, it would be interesting to investigate whether smaller and cost-efficient models can achieve comparable performance \cite{palma2025leveraging}. Specifically, locally deployable models could address privacy concerns that potentially prohibit organizations from transferring log data to external LLM providers \cite{habibzadeh2025large}. We also encourage future studies to compare LLM-based log interpretation with human expert analysis under controlled conditions using identical prompts and data. Such comparisons would provide a more grounded assessment of the practical utility of LLMs in security operations.

\section{Conclusion} \label{conclusion}

In this paper, we introduced CAM-LDS, a publicly available data set of cyber attack manifestations in system log data and security alerts. The data set comprises seven coherent attack scenarios covering 81 distinct MITRE ATT\&CK techniques across 13 tactics and is specifically designed to support reproducible research in LLM-driven interpretation. By focusing on the direct consequences of attack activity in a controlled and reproducible environment, CAM-LDS enables systematic investigation of how adversarial behavior manifests in Linux-centric system logs.

Our analysis revealed that attack manifestations vary substantially in their characteristics. While many attack steps leave explicit command-level traces in logs, others produce indirect indicators such as anomalous event frequencies, changes in system performance metrics, or intrusion detection alerts. At the same time, a considerable fraction of attack steps remain undetected by signature-based IDS, highlighting the limitations of rule-based detection approaches and the importance of complementary analysis methods such as LLMs.

To illustrate the utility of CAM-LDS, we conducted a zero-shot evaluation of LLM-based log interpretation. The results show that LLMs are capable of correctly identifying the underlying attack technique for a substantial portion of attack steps using raw log data alone. Classification performance appears connected to manifestation characteristics; in particular, observable attack commands, high log frequencies, and the presence of intrusion alerts have positive impacts on classification accuracy.

We release CAM-LDS together with the open-source infrastructure, attack scripts, and experimental artifacts to facilitate transparent, reproducible, and comparable research in the area of log-based security analysis. In particular, we encourage future work to incorporate additional contextual and asset information, as well as knowledge from previously classified attack steps, to further improve classification accuracy and more reliably distinguish malicious activity from benign behavior.








\section*{Acknowledgments}
Funded by the European Union under the Horizon Europe Research and Innovation programme (GA no. 101168144 - MIRANDA). Views and opinions expressed are however those of the author(s) only and do not necessarily reflect those of the European Union or the European Commission. Neither the European Union nor the granting authority can be held responsible for them.


\bibliographystyle{elsarticle-num}
\bibliography{mybibfile}

\end{document}